\documentclass[twocolumn]{aa}
\usepackage{graphicx}
\usepackage{longtable,lscape}
\usepackage{txfonts}
\usepackage{natbib}
\bibpunct{(}{)}{;}{a}{}{,}
\begin{document}
\title{Rotation measure synthesis at the 2~m wavelength of the FAN region: \\ Unveiling screens and bubbles.}

   \subtitle{}

   \author{M. Iacobelli\inst{1}\inst{2}, M. Haverkorn\inst{3}\inst{1}, \and P. Katgert\inst{1}}

\offprints{M. Iacobelli; iacobelli@strw.leidenuniv.nl}

\institute{Leiden Observatory, Leiden University, PO Box 9513, 2300 RA Leiden, the Netherlands
\and ASTRON, the Netherlands Institute for Radio Astronomy, Postbus 2, 7990AA, Dwingeloo, The Netherlands
\and Radboud University Nijmegen, Heijendaalseweg 135, 6525 AJ Nijmegen, the Netherlands}

\date{Accepted 04 October 2012}
\authorrunning{Iacobelli et al.}
\titlerunning{Rotation Measure Synthesis at 2~m wavelength of the Fan region: Unveiling screens and bubbles.}

\abstract{
Rotation Measure synthesis of the Westerbork Synthesis Radio Telescope (WSRT) observations at $\lambda \sim 2$~m of the FAN region at \textit{l}=137$^{\circ}$, \textit{b}=+7$^{\circ}$ shows the morphology of structures in the ionized interstellar medium.
}{
We interpret the diffuse polarized synchrotron emission in terms of coherent structures in the interstellar medium and the properties of the interstellar magnetic field.
}{
  We performed statistical analysis of the polarization data cube obtained through Rotation Measure synthesis.  For the first time, cross-correlation is applied to identify and characterize polarized structures in Faraday depth space. Complementary information about the medium are derived from H$\alpha$ emission, properties of nearby pulsars, and optical polarized starlight measurements.
}{
  We find an overall asymmetric Faraday dispersion function in a Faraday depth range of [-13,+5]~rad~m$^{-2}$, which is peaked around $-1$~rad~m$^{-2}$. Three morphological patterns are recognized, showing structures on scales from degrees down to the beam size. The first structure is a nearby synchrotron emission component with low Faraday depth, filling the entire field of view. The second pattern is a circular polarization structure with enhanced (negative) Faraday depth, which has the same morphology as a low-emission region within the third component. This third component is interpreted as the background in which the circular structure is embedded. At low Faraday depth values, a low gradient across the imaged field is detected, almost aligned with the Galactic plane. Power spectra of polarized structures in Faraday depth space provide evidence of turbulence.
}{
A sign reversal in Faraday depth from the nearby component to the circular component indicates a reversal of the magnetic field component along the line of sight, from towards the observer and nearby to away from the observer at large distances. The distance to the nearby, extended component is estimated as $\lesssim 100$~pc, which suggests that this structure corresponds to the Local Bubble wall. 
For the circular component, various physical interpretations are discussed. The most likely explanation is that the circular component seems to be the presence of a nearby ($\sim200$~pc away) relic Str\"{o}mgren sphere, associated with an old unidentified white dwarf star and expanding in a low-density environment.
}
\maketitle
%

\section{Introduction}

The study of the magneto-ionic properties of the interstellar medium (ISM) by multi-frequency polarimetry started in 1960 with the Dwingeloo radio telescope \citep{Westerhout62}. A breakthrough came much later when \citet{Wieringa93} acknowledged the importance of small angular scale structures (down to a few arcmins) in polarized intensity (PI) or polarization angle with no detection in total intensity. Such a fine structure was explained in terms of Faraday rotation of the diffuse, linearly polarized, Galactic synchrotron radiation background by a highly structured foreground magneto-ionic medium. The cosmic rays propagating through the Galactic magnetic field are responsible for the radio synchrotron emission, and the magneto-ionic medium is primarily the warm ionized ISM. As a result polarimetric observations of the Galaxy provide insight into the ionized ISM and the Galactic magnetic field.

The FAN region is a highly polarized, extended region in the second Galactic quadrant at low Galactic latitudes. Its polarization properties were first investigated by \citet{Bingham and Shakeshaft67}, who discovered a circular structure in the rotation measure (RM) map. From higher angular resolution observations, \citet{Verschuur69} suggested a connection of the observed structure with the B2\,IV:e star HIP\,15520 that is located close to the centre of the circular structure. 

\citet{Haverkorn03} presents multi-frequency polarimetry radio observations at five wavelengths in the range of $\lambda = 0.80-0.88$~m. They point out a ring structure in PI along with an extended pattern of depolarization canals in maps of polarized intensity. They constructed an RM map from these five wavelengths and interpret the ring as a purely magnetic flux tube.

\citet{Bernardi09} observed the FAN region at $\lambda = 1.93-2.16$~m in order to characterize the properties of the foreground for epoch of re-ionization experiments. They detected total intensity and polarization fluctuations for the first time in the Galactic diffuse foreground emission at this wavelength. Rotation measure synthesis (RM-synthesis) was used to image this region in the sky revealing a complex and structured distribution of polarized signals.

In this paper, we re-analyse the data from Bernardi et al.\ to study the diffuse Galactic polarized emission. In Section 2 we present the data and illustrate the working principle of RM synthesis technique. In Section 3 we describe the main properties of total and polarized intensity. In Section 4 the polarized data in Faraday depth space are discussed. In Section 5 we describe the main components and their properties. In Section 6 we present a simple model to explain the observed structures and their features. Finally in Section 7 we summarize and conclude.

\section{Data analysis}
\label{s:data}

The data were collected with the east-west Westerbork Synthesis Radio Telescope (WSRT) array in 2007. Baselines between 36~m and 2.7~km were used to obtain good \textit{uv}-coverage. Data reduction was performed by Gianni Bernardi using the AIPS++ package, and for a description of their reduction we refer to \citet{Bernardi09}. The corresponding main properties are given in Table~1. Ionospheric propagation effects are direction- and time-dependent and affect the phase of the signal received by the interferometer. They also affects the polarization, giving rise to time-variable Faraday rotation. Moreover, differential Faraday rotation by the ionosphere will alter Stokes Q and U, which could lead to depolarization but could also lead to spurious extra polarization, thus affecting the results of the polarization imaging. Over the six days of observation the mean total electron content (TEC) value (provided by the Center for Orbit Determination in Europe (CODE)) and the mean intensity of the geomagnetic field (calculated using the International Geomagnetic Reference Field (IGRF10)) give us a typical $RM_{ion}\lesssim 0.3$~rad~m$^{-2}$, which implies a phase shift of about 80 degrees at the lower observed frequency. The spread of the $RM_{ion}$ over days of observation is 0.1~rad~m$^{-2}$. Since the resolution in Faraday depth is about 3~rad~m$^{2}$ (see Table~1), ionospheric Faraday rotation will hardly change the analysis or conclusions, so that corrections for the ionospheric Faraday rotation were not applied. \\
To enhance sensitivity for diffuse emission, the resolution of the Stokes Q and U maps was decreased. In this paper we use the full angular resolution Stokes I map (i.e. $\Delta\theta_{0} = 2$~arcmin) and Stokes Q,U maps with angular resolution $\Delta\theta = 4.2$~arcmin. No correction was performed for the primary beam attenuation in order to preserve a uniform noise level across the images. In the following section, we summarize the principle of the RM-synthesis technique. 

\begin{table*}
\label{table:1}
\centering
\caption{Observational set up and basic properties of the RM-synthesis cube, from \citet{Bernardi09}}
\begin{tabular}{lll}
\hline\hline
Data\\
\hline
\textit{l},\textit{b} & 137$^{\circ}$  +7$^{\circ}$ & Galactic coordinates\\
N$_{bands}$ & 8 & Number of spectral bands\\
$\nu_{0,band}$ & 139.3, 141.5, 143.7, 145.9, & \\ 
 & 148.1, 150.3, 152.5, 154.7~MHz & Central frequency of each band\\   
$\Delta\nu_{0,band}$  & 2.5~MHz & Width of each band\\
$\delta\nu_{ch}$ & 9.8 kHz & Frequency resolution\\
$\Delta \theta_0$ & 2$\arcmin$ x 2.2$\arcmin$ & Beam size\\
\hline\hline
($\alpha,\delta,\phi$)-cube\\
\hline
$\Delta\phi$ & 3 rad~m$^{-2}$ & RMSF width\\
$\delta\phi_{scale}$ & 0.85 rad m$^{-2}$ & Max scale in $\phi$\\
$\delta \theta$ & 4.2$\arcmin$ & Beam size\\
$\xi$ & 1 mJy beam$^{-1}$ = 1 K & Conversion factor\\
\hline
\end{tabular}
\end{table*}

\subsection{RM-synthesis}
\label{s:rmsynth}

Low frequency polarimetric studies of the diffuse Galactic radio emission are a valuable tool to investigate the ionized and magnetized components of the interstellar medium. Generally, these studies have been based on RM measurements calculated as the slope of a linear fit to the polarization angle as a function of the wavelength squared. However, a new analysis technique has recently been applied, the so-called RM-synthesis \citep{Burn66}. This technique has only recently become practically applicable due to technical and computational advances, and was developed and used for the first time by \citet{Brentjens05}. Its main advantages with respect to the standard RM technique are that it provides a mapping of the linearly polarized emission as a function of Faraday depth ($\phi$), the Faraday dispersion function ($F(\phi)$), and that bandwidth depolarization and depth depolarization are much less severe. RM-synthesis is implemented as a weighted (by the sampling function $W(\lambda^{2})$) Fourier transform between $F(\phi)$ and the measured complex polarization vector $P(\lambda^{2})$:
\begin{equation} P(\lambda^{2}) = W(\lambda^{2}) \int_{- \infty}^{+ \infty} F(\phi) e^{2i\phi\lambda^{2}} d\phi \: ,\end{equation}
with the Faraday depth \begin{equation} \phi = 0.81 \int_{d_{1}[pc]}^{d_{2}[pc]} n_\mathrm{e} \vec{B} \cdot d\vec{l} \: \end{equation} where the electron density $n_{e}$ and the magnetic field $\vec{B}$ are given in cm$^{-3}$ and $\mu$G, respectively. 
Since we can only sample a finite positive range of wavelengths, the inversion of the Fourier transform is incomplete, and we only get an approximation of $F(\phi)$. However, three physical quantities directly linked to the experimental set up (the channel width $\delta\lambda^{2}$, the width of the $\lambda^{2}$ distribution $\Delta\lambda^{2}$, and the shortest wavelength squared $\lambda_{min}^{2}$) can be used to characterize the detection capability: 
\begin{itemize}
\item the maximum detectable Faraday depth, constrained by the channel width $\delta\lambda$: $\phi_{max} \approx \sqrt{3}/\delta\lambda^2$;
\item the maximum scale detectable in Faraday depth to which the sensitivity is reduced to 50\%, which is constrained by the lowest observed wavelength $\lambda_{min}$: $\delta\phi_{scale} \approx \pi / \lambda_{min}^{2}$;
\item the resolution in Faraday depth space defined as the half power width of the rotation measure spread function\footnote{The point spread function in Faraday depth space is generally called the rotation measure spread function.} (RMSF), constrained by the total wavelength range $\Delta\lambda^2$: $\Delta\phi \approx 2\sqrt{3} / \Delta\lambda^{2}$). This defines the minimum separation between separate synchrotron emitting structures that is detectable.
\end{itemize}
For our frequency coverage, $\phi_{max} \approx 2650$~rad~m$^{-2}$, $\delta\phi_{scale} \approx 0.85$~rad~m$^{-2}$, and $\Delta\phi \approx 3$~rad~m$^{-2}$. With these data we can therefore only detect Faraday thin structures, since the resolution is greater than the maximum detectable scale. Since the polarized radiation is a vector quantity, astrophysical information is stored in both intensity and angle. However, due to the generally low signal-to-noise ratio (S/N) per resolution element, the solution of the $n\pi$-ambiguity is challenging so we do not use polarization angle maps for a quantitative derivation, but only to infer some basic properties. For this reason, we focus on polarized intensity maps and maps of Stokes~U. Maps of Stokes~Q show similar characteristics to Stokes~U.

\subsection{RM-synthesis data cubes}

Maps for Stokes Q,U, and PI were obtained in Faraday depth space by applying the RM-synthesis technique to the 4.2~arcmin Stokes Q,U maps in the frequency domain. No bias correction was applied to the polarized intensity maps but owing to the high side lobe level of the RMSF, an RM CLEAN \citep{Heald09} was performed with a threshold of 5~mJy. The polarized emission from our Galaxy is expected at low Faraday depth values $|\phi| \lesssim 10^{2}$~rad~m$^{-2}$ \citep[see][]{Clegg92}. Inspection of this range of Faraday depth reveals the presence of significant polarized emission only within a few tens of frames around $\phi = 0$~rad~m$^{-2}$ \citet{Bernardi09}. The PI content of the data out of this range of Faraday depths has not been checked. For the data analysis, we limited the Faraday depth range of this data set to $[-50,+50]$~rad~m$^{-2}$.

\section{Observational results}

Two well known basic properties of the diffuse polarized emission \citep[see][]{Wieringa93} are visible in the  images (see e.g.\ Figs.~\ref{instrumental_features} and~\ref{rgb_image}): 
\begin{itemize}
\item the intensity attenuation due to the WSRT primary beam (about 2.5$^{\circ}$ at 150~MHz). This is a strong argument for excluding an instrumental origin of the polarization fluctuations, and it suggests a wider extent on the sky of the detected patterns of polarized emission;
\item the absence of a correlation between total intensity and polarized intensity, which indicates that the polarization fluctuations are caused by Faraday rotation of the diffuse synchrotron emission background by the intervening magneto-ionized medium.
\end{itemize}
In the following sections, we describe the global properties of the signals detected in polarized intensity and total intensity.

\subsection{Total intensity}

\begin{figure}
\resizebox{1.0\columnwidth}{!}{\includegraphics{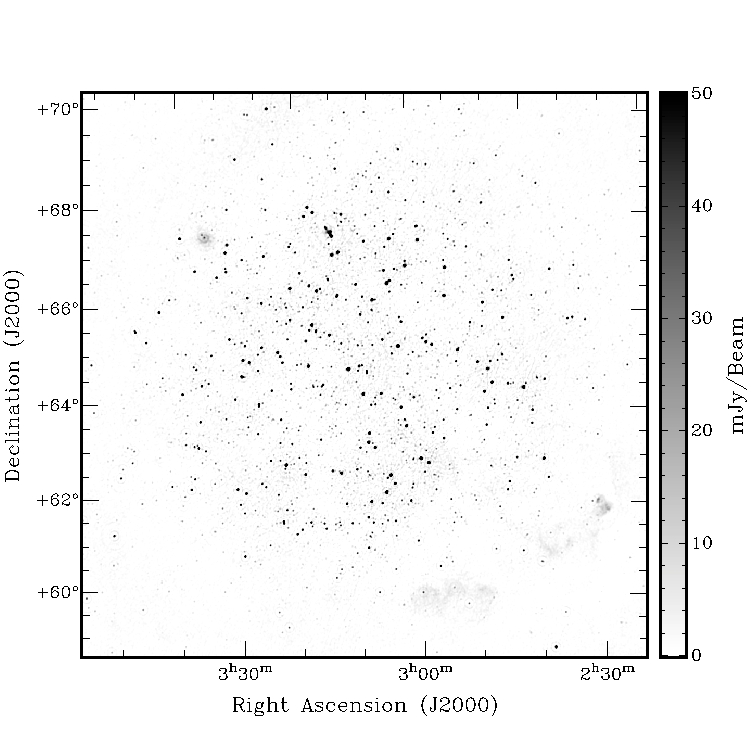}}
\caption{\label{stoke_I} \textit{Total intensity map at 150~MHz with an angular resolution of $\simeq 2\arcmin \times 2.2\arcmin$. The diffuse structures towards the upper left and lower right corner are the galaxy IC\,342 and the \ion{H}{ii} region complex W3/W4/W5 respectively.}}
\end{figure}

The Stokes I map displayed in Fig.~\ref{stoke_I} has a size of about 12$^{\circ}$ in declination and 12$^{\circ}$ in right ascension, and its dynamic range is about 1450:1 at an angular resolution of $2' \times 2.2'$ FWHM. The only extended Galactic object detected is the \ion{H}{ii} region complex W3/W4/W5 in the south-west corner, approximately from $(\alpha,\delta)_{J2000} \approx (02^{h}25^{m},62^{\circ}.0)$ to $(\alpha,\delta)_{J2000} \approx (02^{h}53^{m},60^{\circ}.4)$. Faint Stokes I emission is also seen from the spiral galaxy IC\,342 at $(\alpha,\delta)_{J2000} \approx (03^{h}47^{m},68^{\circ}.0)$ and from the giant double radio galaxy WNB\,0313+683 at $(\alpha,\delta)_{J2000} \approx (03^{h}18^{m},68^{\circ}.3)$. No large-scale diffuse mission in the primary beam is seen at 150~MHz because the interferometer filters away power on the largest scales. However, \citet{Bernardi09} detected extended faint, mottled, small-scale total intensity emission from the Galactic foreground. From the power spectrum these authors suggest a minimum spatial scale of $\sim$12~arcmin for these Stokes~I fluctuations.

\subsection{Polarized intensity}

\begin{figure}
\resizebox{1.0\columnwidth}{!}{\includegraphics{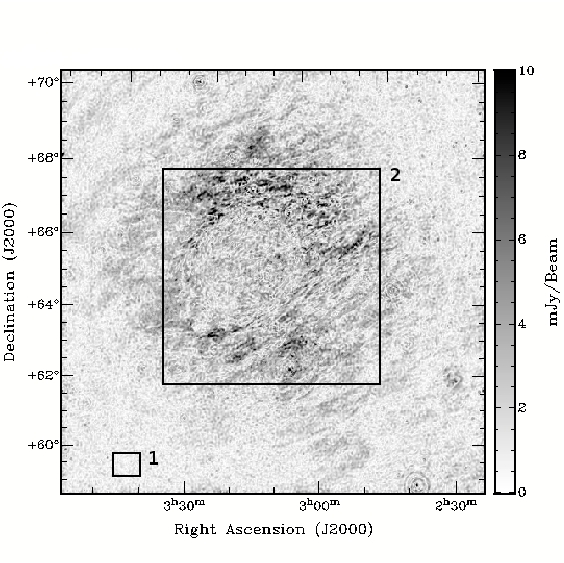}}
\caption{\label{instrumental_features} \textit{Frame of the RM-synthesis cube at Faraday depth $\phi = -1$~rad~m$^{-2}$ displaying the extended polarized emission. Grating rings can be seen around the bright  point sources at $(\alpha,\delta)_{J2000} \approx (02^{h}38^{m},59^{\circ}.1)$ (close to the edge of the image), and $(\alpha,\delta)_{J2000} \approx (03^{h}06^{m},62^{\circ}.5)$. The drawn boxes are chosen to infer the noise behaviour (box \textbf{1}) and derive the cross-correlation profile (box \textbf{2}), see text.}}
\end{figure}

Figure~\ref{instrumental_features} shows the polarized intensity at a Faraday depth $\phi = -1$~rad~m$^{-2}$. Grating rings are seen around brighter compact sources, but owing to their multiplicative nature, such error patterns are only visible around bright Stokes~I sources, so their effect is negligible in the map except in the direct vicinity of bright point sources. Unwanted instrumental signals are visible in the Stokes Q, U, and PI maps as elongated stripes and/or waves crossing the field. As pointed out by \citet{deBruyn05} and \citet{Pizzo11}, the WSRT off axis polarization has a strong frequency dependence with a period of 17~MHz, which causes peaks in the Faraday spectrum at about 42~rad~m$^{-2}$. At our frequency range, this 17~MHz ripple appears as a broad peak at the edges of the selected Faraday depth range (see Fig.~\ref{histo_profile}). These are due to sources Cas~A and Cyg~A entering the telescope side lobes.

\begin{figure}
\resizebox{1.0\columnwidth}{!}{\includegraphics{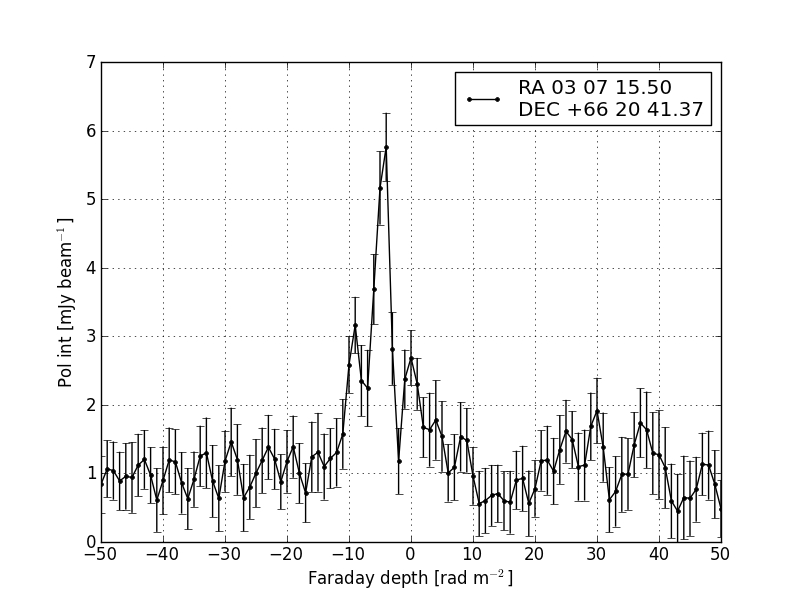}}
\caption{\label{rm_profile} \textit{Example of Faraday dispersion spectrum with emission at a single Faraday depth along the line of sight, averaged over a beam-sized box region. The noise level is indicated by $1\sigma$ error bars.}}
\end{figure}

\begin{figure}
\resizebox{1.0\columnwidth}{!}{\includegraphics{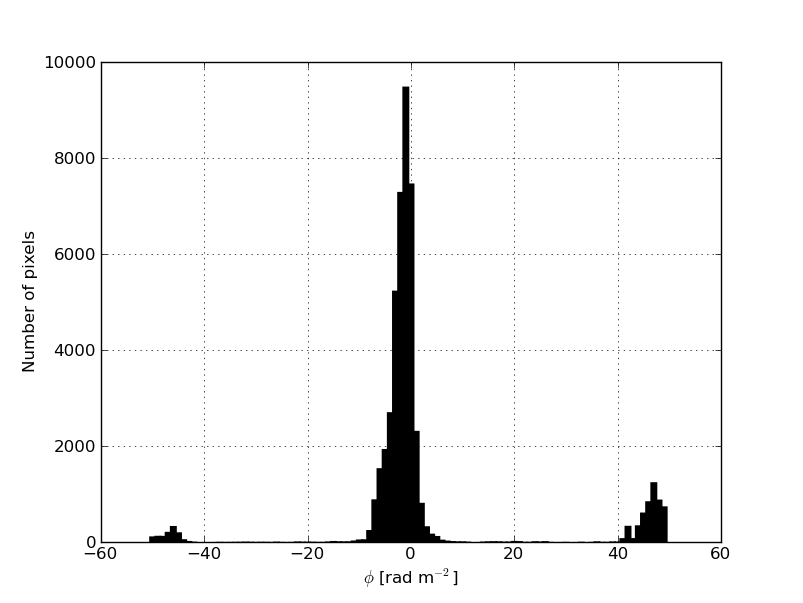}}
\caption{\label{histo_profile} \textit{Distribution of  polarized intensity as a function of Faraday depth, above a threshold of 5~mJy. At $|\phi| \geq 40$~rad~m$^{-2}$ the residual signal from side lobes of Cyg A and Cas A are visible, while the Galactic emission occurs around $-1$~rad~m$^{-2}$ and has an asymmetric profile.}}
\end{figure}

The overall properties of the polarized signal distribution as a function of Faraday depth are shown in Figure~\ref{histo_profile} and an example of Faraday dispersion spectrum is displayed in Figure~\ref{rm_profile}. The distribution of polarized signal includes the whole field, i.e.\ $480 \times 480$ lines of sight. We counted at each Faraday depth the number of pixels having a polarized intensity above a 5~mJy~beam$^{-1}$ threshold ($\sim 5$ times the PI mean noise level). Polarized emission is detected over a wide range from positive to negative $\phi$ values. Two main features are clearly seen:
\begin{itemize}
\item a central peak at slightly negative $\phi$ where the most of polarized emission is,
\item two wings towards the edges of our $\phi$ range, due to the residuals from Cas~A and Cyg~A.
\end{itemize}
Figure~\ref{histo_profile} makes it evident that the Galactic emission from this region is concentrated in a very narrow $\phi$ range from about $-$10 to +5~rad~m$^{-2}$. The main central feature in the distribution is clearly asymmetric; we interpret this as an evidence of a multi-component signal and we discuss this in Sects. 5 and 7.

\section{The polarization cube in Faraday depth space}

In this section we discuss the noise properties and the connection between RMSF width and cross-correlation in Faraday space.

\subsection{Noise properties and errors}

Since we are dealing with the measurement of a pseudo-vector - i.e. the polarization - in the presence of random noise we look at noise behaviour. While the two Stokes Q and U parameters are Gaussian distributed, the linearly polarized intensity \begin{equation} PI = \sqrt{Q^{2} + U^{2}} \: \end{equation} follows an asymmetric positive definite distribution: the Rice distribution, which is dependent on signal-to-noise and responsible for a bias towards too high values of PI at low S/N \citet{Wardle74}. \\

We tested the Gaussian behaviour of Stokes~Q and U maps and estimated the noise in PI as a function of Faraday depth under the assumption of a Rayleigh distribution of polarized intensities. The noise was determined from a small region assumed to contain no emission from the sky, so that all detected emission is noise (see box \textbf{1} in figure~\ref{instrumental_features}). We assume uniformity of the noise across the field. Since in every Faraday depth frame we detect emission (either from the ISM or from grating lobes from strong out-of-field sources), this uniformity could not be tested. Noise values in Stokes~Q and U are similar; also, the assumption of Gaussian noise in Stokes components Q and U is satisfied, because the standard deviation of the sample and the width of fitted normal distribution agree with each other.

\begin{figure}
\resizebox{1.0\columnwidth}{!}{\includegraphics{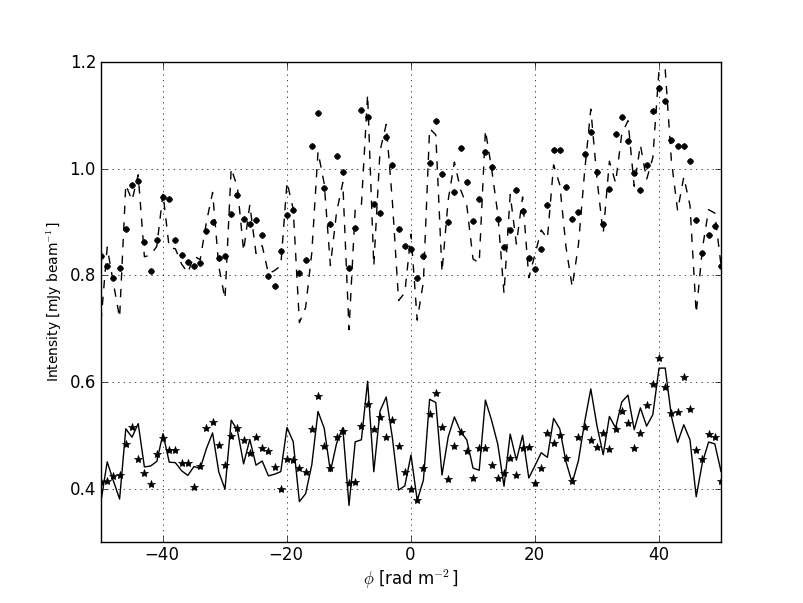}}
\caption{\label{noise_profile} \textit{Noise behaviour in the PI map derived from a region denoted by  box \textbf{1} in figure~\ref{instrumental_features} as a function of Faraday depth under the assumption of Rayleigh distribution statistics. The expected mean and variance values of the noise are indicated by the dashed and solid lines, while the measured values indicated by circles and stars refer to the noise mean and variance, respectively.}}
\end{figure}

The noise properties of PI as a function of the Faraday depth under the assumption of Rayleigh distribution statistics are shown in figure~\ref{noise_profile}. Good agreement is found for the expected (i.e. $\sigma_{PI} = 0.66 \sigma_{U}$) and the measured noise variance, as well as for the expected (i.e. $<\mu_{PI}> = 1.25 \sigma_{U}$) and the measured noise mean value. A mean noise level of $\simeq 1$~mJy~beam$^{-1}$ is found in PI over a $\phi$ range of $[-50,+25]$~rad~m$^{-2}$ while an increase in the mean and variance values is seen in the Faraday depth range above $+25$~rad~m$^{-2}$ due to the presence of residual signal from side lobes of Cyg~A and Cas~A. Because the Rayleigh statistics a clipping of PI values at $3\sigma$ allows to mitigate the impact of Ricean bias below 10\% level. Since our analysis does not depend on exact values of polarized intensity, but rather uses rough estimates of ISM parameters, bias corrections were not applied. \\

No corrections for the off-axis instrumental polarization were applied. We estimated its magnitude and radial dependence from the fractional polarization values at $\phi =$ 0~rad~m$^{-2}$, where instrumental polarization is expected. We found a mean level of $\simeq 2$\% within 3$^{\circ}$ (primary beam), which rises to $\simeq 9$\% at the edges of the field of view. There are not enough sources left in the maps after the subtraction to derive the angular dependence of the antenna pattern.

In analogy to radio interferometric observations, the standard errors in Faraday depth measurements can be obtained in an RM-synthesis cube as: $\sigma_{\phi} = 0.5 \Delta\phi/(S/N) \sim 1.5/(S/N)$~rad~m$^{-2}$, where $\Delta\phi$ is the RMSF width and \textit{S/N} the PI signal-to-noise ratio. Because we mostly deal with spatially extended emission over several ranges of Faraday depths, we prefer to provide a general upper limit for the $\phi$ uncertainty at each frame as set by the mean \textit{S/N} of the emission across the maps. Because the chosen $3\sigma$ clip level (i.e. $S/N\gtrsim3.0$): $\sigma_{\phi} \lesssim 0.50$~rad~m$^{-2}$. Therefore the total Faraday depth estimation error in our 2~m data consists of two main contributions: the systematic error due to the ionospheric Faraday rotation (as discussed in Section \ref{s:data}) and the standard error (as shown in Figs.~\ref{rm_profile} and \ref{faraday_dispersion_spectra}), having comparable amplitudes.

\subsection{Cross-correlation in $\phi$ space}

In RM-synthesis the standard estimator for the nominal resolution in Faraday space is the RMSF width. However, as in aperture synthesis, there is some information present on scales smaller than the beam size, as can be seen in figure~\ref{stoke_U}. This figure shows polarized intensity changes between neighbouring frames (i.e. $\Delta\phi = 1$~rad~m$^{-2}$), about one third of the RMSF width. 
An effective approach to investigating $\phi$-resolution is cross-correlation. Cross-correlation between frames can be used as a probe to infer the presence of structures in the data and their Faraday thickness. We calculate the cross-correlation coefficient for the frames $a$ and $b$ with a frame gap $\delta\phi$ ($C_{\delta\phi}$) as \begin{equation} C_{\delta\phi} = \frac{1}{N-1} \sum_{i,j} \frac{(a_{ij} - \overline{a}) (b_{ij} - \overline{b})}{\sigma_{a}\sigma_{b}} \: ,\end{equation} where $N$ is the total number of pixels, ($\overline{a},\overline{b}$) are the mean values of the frame, and ($\sigma_{a},\sigma_{b}$) are their standard deviations. The frame gap $\delta\phi = |a-b|$ is the distance between frames $a$ and $b$ in rad~m$^{-2}$.

\begin{figure}
\resizebox{0.97\columnwidth}{!}{\includegraphics{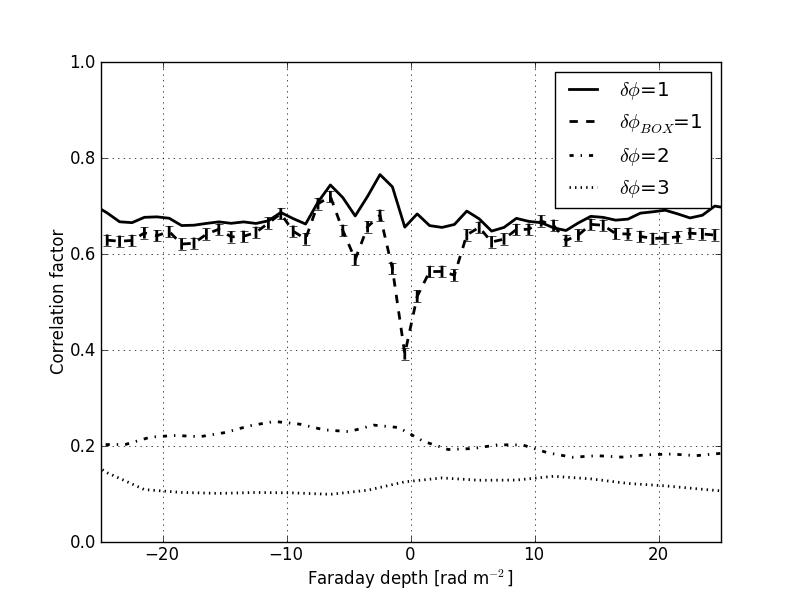}}
\caption{\label{corr_all_field} \textit{Cross-correlation coefficients as function of Faraday depth for different frame gaps $\delta\phi=1$~rad~m$^{-2}$ (solid), $\delta\phi=2$~rad~m$^{-2}$ (dash-dotted), and $\delta\phi=3$~rad~m$^{-2}$ (dotted). The cross-correlation at $\delta\phi=1$~rad~m$^{-2}$ for only the emission within box \textbf{2} in figure~\ref{instrumental_features} is shown by the dashed line. The uncertainties are shown by $1\sigma$ error bars.}}
\end{figure}

We studied the correlation length in Faraday depth space of the PI data by considering the whole field of view and by focussing on a square region in the centre of the field (given by box \textbf{2} in figure~\ref{instrumental_features}). In figure~\ref{corr_all_field} we show the cross-correlation coefficients for different frame gaps ($\delta\phi$).

For a fixed convolving RMSF, the reference level for data correlation/anti-correlation is shifted from zero to a certain positive value. This offset is due to the convolution of the noise of each frame with the RMSF, which introduces a degree of correlation of the noise in the frames. The cross-correlation of frame gaps $\delta\phi=1,2,3$~rad~m$^{-2}$ for noise-dominated frames convolved with an RMSF of 3~rad~m$^{-2}$ is approximately 0.68, 0.22 and 0.1, respectively. Therefore the average levels in Fig.~\ref{corr_all_field} correspond to a convolving function with a width of $\simeq 2.95$~rad~m$^{-2}$, in good agreement with the width of the RMSF. This means that outside the range of Galactic signal of the Faraday depth, from about $-$10 to +5~rad~m$^{-2}$, the cross-correlation profile is consistent with a pure noise signal. Therefore for $\delta\phi=1$~rad~m$^{-2}$, a correlation coefficient above 0.68 indicates a positive correlation between two frames, a correlation coefficient of about 0.68 indicates no correlation while a correlation coefficient below 0.68 indicates an anti-correlation.

Several features in the cross-correlation profile for $\delta\phi=1$ can be seen, which disappear as the frame gap increases. Therefore they are not broader than the RMSF width, which means these structures are Faraday thin.  For a frame gap of $\delta\phi = 1$~rad~m$^{-2}$, prominent peaks are seen at $\phi \approx -5$~rad~m$^{-2}$ and $\phi \approx -2$~rad~m$^{-2}$. Three more tentative, small peaks, possibly related to fainter extended emission, are observed around $\phi \approx -10$~rad~m$^{-2}$, $\phi \approx +1$~rad~m$^{-2}$, $\phi \approx +5$~rad~m$^{-2}$. 

Calculating the cross-correlation coefficients with the same frame gap of $\delta\phi = 1$~rad~m$^{-2}$ in the smaller box indicated as \textbf{2} in Figure~\ref{instrumental_features} shows interesting results. Correlated peaks at $\phi \approx -10,-6,-2$~rad~m$^{-2}$ are still seen, although their amplitudes have changed. However, the most conspicuous changes are seen in the range of Faraday depth $\phi \sim [-2,+4]$~rad~m$^{-2}$ where the cross-correlation profile is turned into anti-correlation, with two possible small peaks at $\phi \approx 0$~rad~m$^{-2}$ and $\phi \approx +2$~rad~m$^{-2}$. The polarized emission in this $\phi$-range is spatially anti-correlated. Simple simulations show that qualitatively this behaviour can be caused by a spatial gradient in Faraday depth space across these frames. The presence of multiple Faraday components within the peak was also confirmed by comparing the width of the peaks in the spectrum with the width of the RMSF. We checked the behaviour of some representative lines of sight through the ``bubble'', ``ring'', and ``curtain'' components. From the fitting of the limited sample of Faraday dispersion spectra, we find clues of multiple poorly separated components, in agreement with the cross correlation result. \begin{itemize}
\item the RM peaks associated with the noise or artefacts (e.g. due to the lobes of the RMSF) show a systematically low $FWHM<2.7$~rad~m$^{-2}$, while RM peaks associated with the three components have $FWHM \sim [2.9,3.2]$rad~m$^{-2}$, in agreement with the expected RMSF width;
\item Some lines of sight show excess of widths or complex morphology (see e.g. spectrum of line of sight 4 in Figure~\ref{faraday_dispersion_spectra}), which make them unsuitable for a single component fitting step. These cases can be found in all three components, but are more evident for the ``curtain'' component;
\item Often spectra also have associated main lobes that are higher than the $\sim$35\% level expected from the RMSF profile. In conclusion we find agreement between the result of this test and the cross-correlation.
\end{itemize}

\section{Definition and description of structures}

The ($\alpha,\delta,\phi$)-cube contains a wealth of structures. In this section we briefly describe the main morphological features in the PI and Stokes U maps as a function of Faraday depth. First, there are spatially compact and isolated structures, such as Galactic and extragalactic Stokes~I sources. These objects primarily ``illuminate'' the intervening Galactic ISM, and reveal its Faraday depth. If their structure is multi-component, then internal Faraday dispersion can be significant and they may actually show emission over a range of Faraday depths. Their definition is based on their compactness in ($\alpha,\delta$) and, occasionally, on association with known objects. The second class of structure is the spatially (very) extended emission detected in polarization. These can be characterized by their properties of PI, Q, or U in Faraday space, in combination with the distribution of polarization angles. \\

\begin{figure*}
\centering
\resizebox{0.9\columnwidth}{!}{\includegraphics{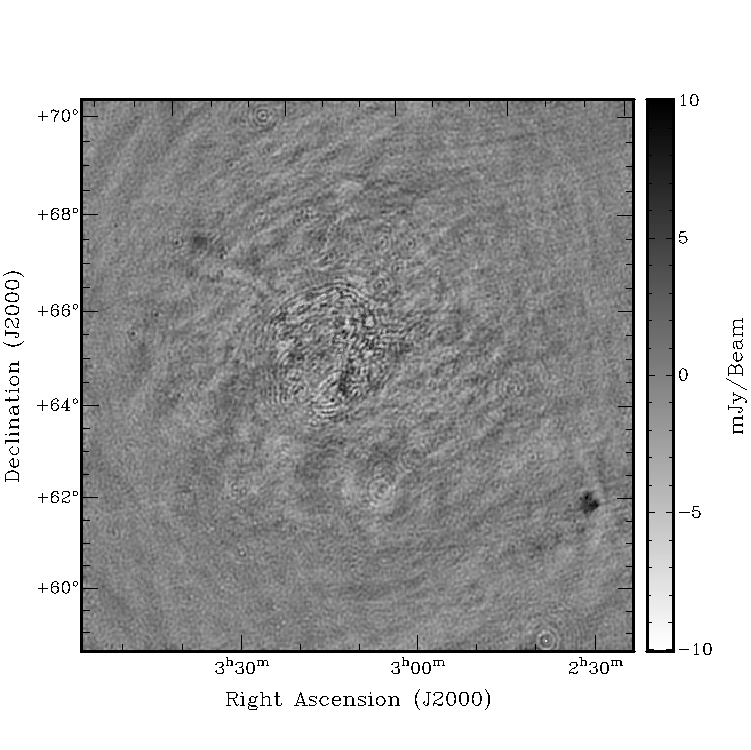}}
\resizebox{0.9\columnwidth}{!}{\includegraphics{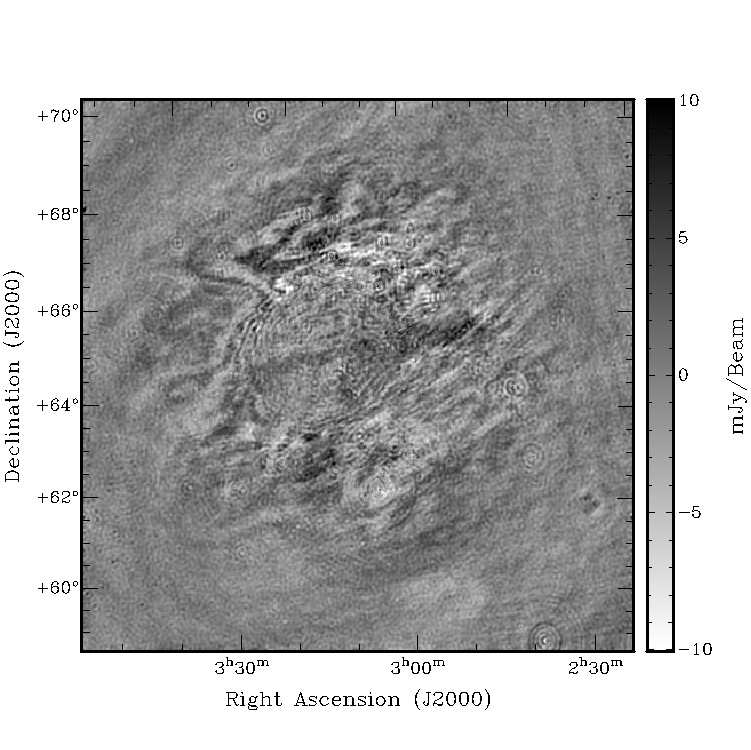}}
\resizebox{0.9\columnwidth}{!}{\includegraphics{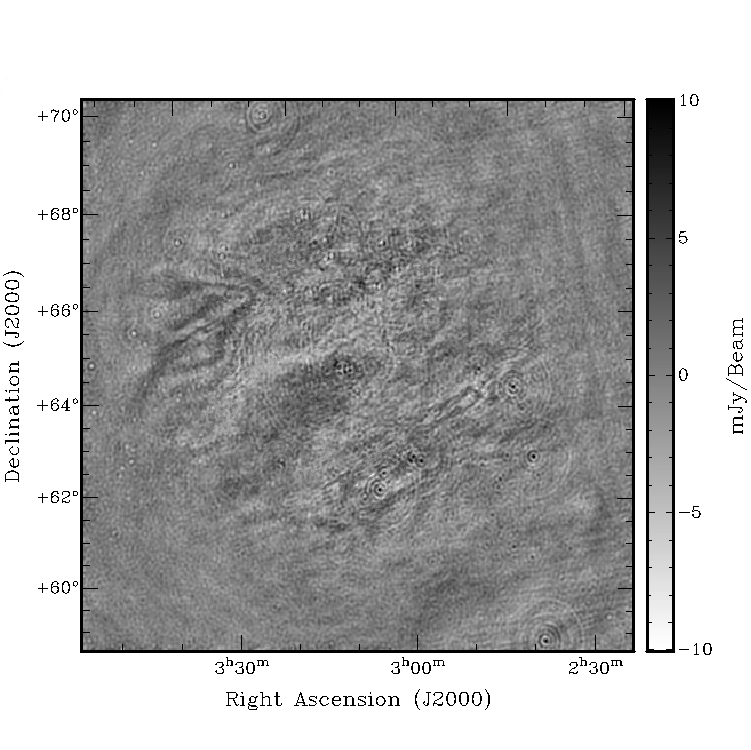}}
\resizebox{0.9\columnwidth}{!}{\includegraphics{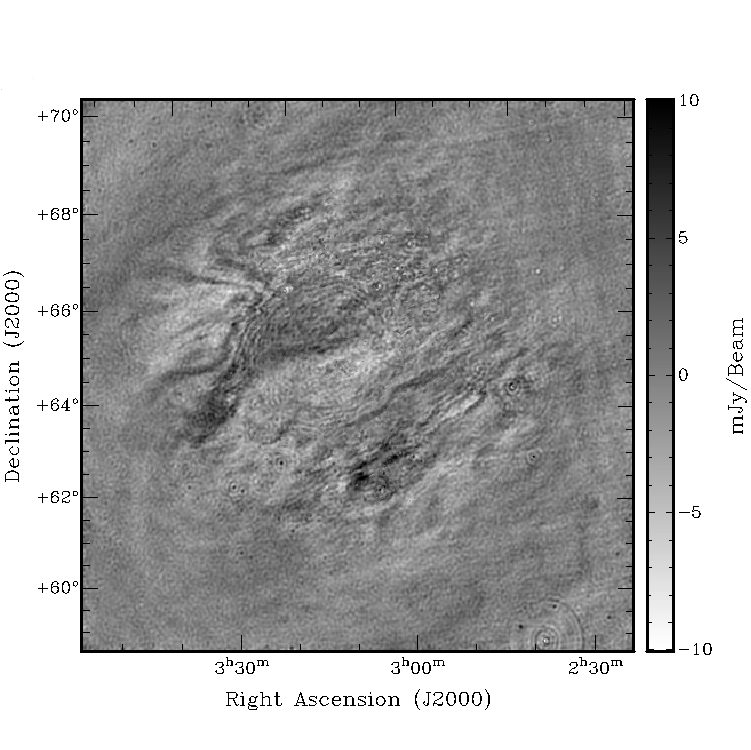}}
\caption{\label{stoke_U} \textit{Stokes U emission at four representative Faraday depth values. The images correspond to $\phi = -5$~rad~m$^{-2}$ (upper left panel), $\phi = -1$~rad~m$^{-2}$ (upper right panel),  $\phi = 0$~rad~m$^{-2}$ (lower left panel) and $\phi = +1]$~rad~m$^{-2}$ (lower right panel).}}
\end{figure*}

We identified various components in the RM-synthesis cube based on morphological consistency and coherence in Faraday depth and name these components the ``bubble'', the ``ring'', and the ``curtain''. These components are shown in polarized intensity in Figure~\ref{rgb_image}.

\begin{itemize}
\item The discrete PI emission from the ``bubble'' appears at $\phi \sim -10$~rad~m$^{-2}$ in the centre of the field of view. A circular filled structure develops and expands with a maximum of polarized intensity at $\phi \sim -5$~rad~m$^{-2}$ and an angular diameter of $\sim$3$^{\circ}$. 
  Also, close to the center a hole in the emission develops and expands until the next structure - the ``ring'' - comes up. The Stokes U map (see Fig.~\ref{stoke_U} upper left panel) depicts a complex pattern of small-scale wrapped structures surrounded by larger and smoother ones.  \item The ``ring'' structure consists of extended emission combined with narrow canals of low polarized intensity around a central, almost circular, edge. The ``ring'' can be considered as a smooth transition from the ``bubble''. Its polarized emission reaches a maximum around $\phi \sim -1$~rad~m$^{-2}$ and shows bright PI emission occurring in the north-western part of the field with longer depolarization canals \citep{Haverkorn00}, with lengths up to a few degrees. The orientation of these canals is mostly parallel to the Galactic plane, in agreement with the observation of \citet{Haverkorn03}. The Stokes (Q and) U maps (see Fig.~\ref{stoke_U}, upper right panel) show a rapid sinusoidal behaviour at the edge of the ``ring'', corresponding to a gradient in polarization angle. 
\item Finally, a significant pattern of faint, extended polarized emission -the ``curtain''- is found at $\phi \sim [0,+5]$~rad~m$^{-2}$. A fast transition is seen around $\phi \approx +2$~rad~m$^{-2}$ from the previous ``ring'' component to the ``curtain''. We interpret this as a spatial gradient of polarized emission in Faraday space around $\phi \sim -2$~rad~m$^{-2}$ and $\phi \sim 0$~rad~m$^{-2}$, which is responsible for the abrupt decrease in correlation.  Stokes U emission (Fig.~\ref{stoke_U} lower panels) reveals extended features on larger scales than the ``bubble'' and ``ring'' components. Also, whereas the ``bubble'' and ``ring'' show morphological correspondences, these are not present for the ``curtain''.
\end{itemize} These three components in the RM-synthesis cube correspond to the peaked features seen in the cross-correlations profiles (Figure~\ref{corr_all_field}) around $\phi \sim -6,-2,+2$~rad~m$^{-2}$, respectively.

\subsection{Diffuse foreground polarized emission}

\begin{figure*}
\centering
\resizebox{0.9\columnwidth}{!}{\includegraphics{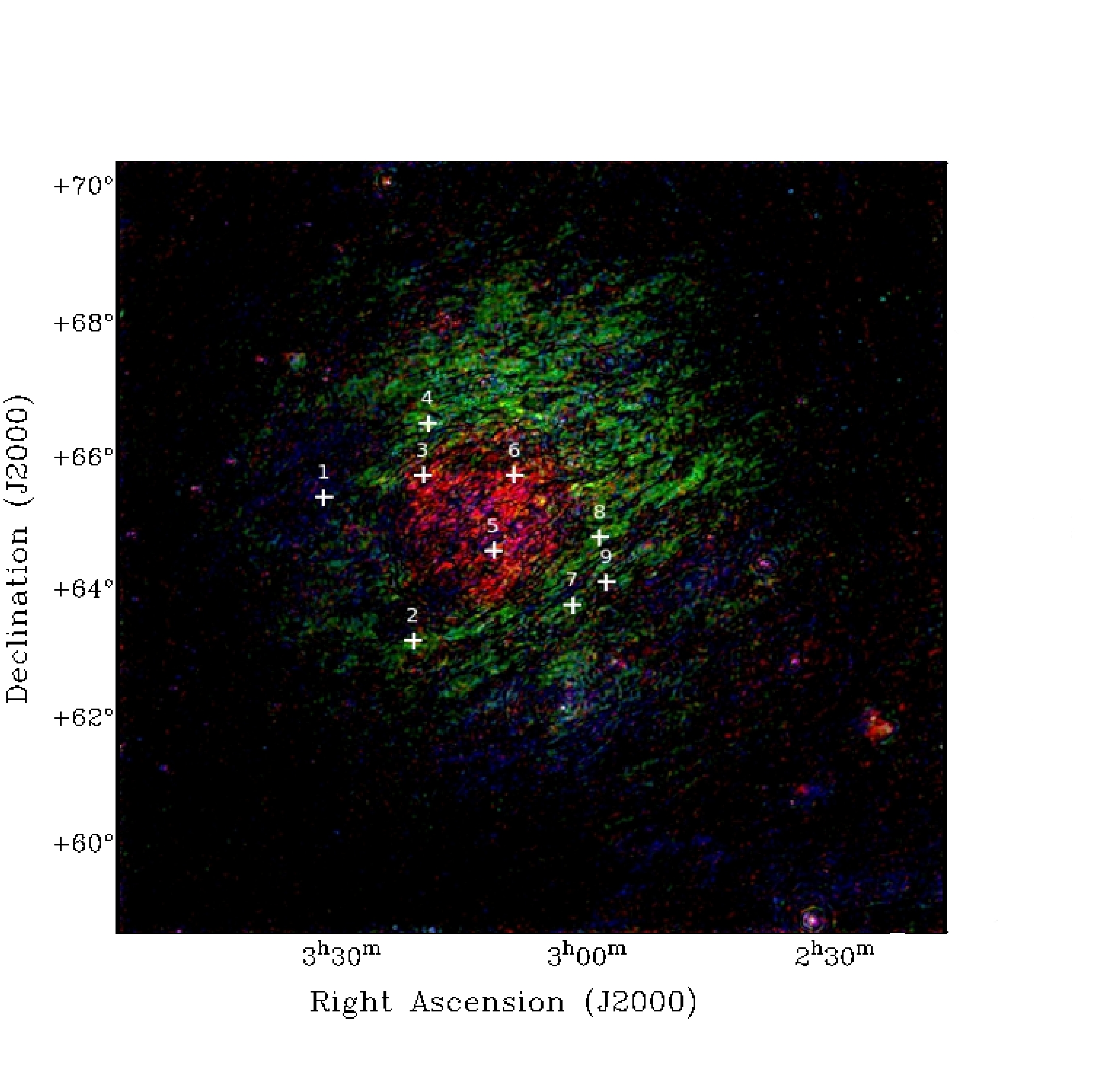}}
\resizebox{0.9\columnwidth}{!}{\includegraphics{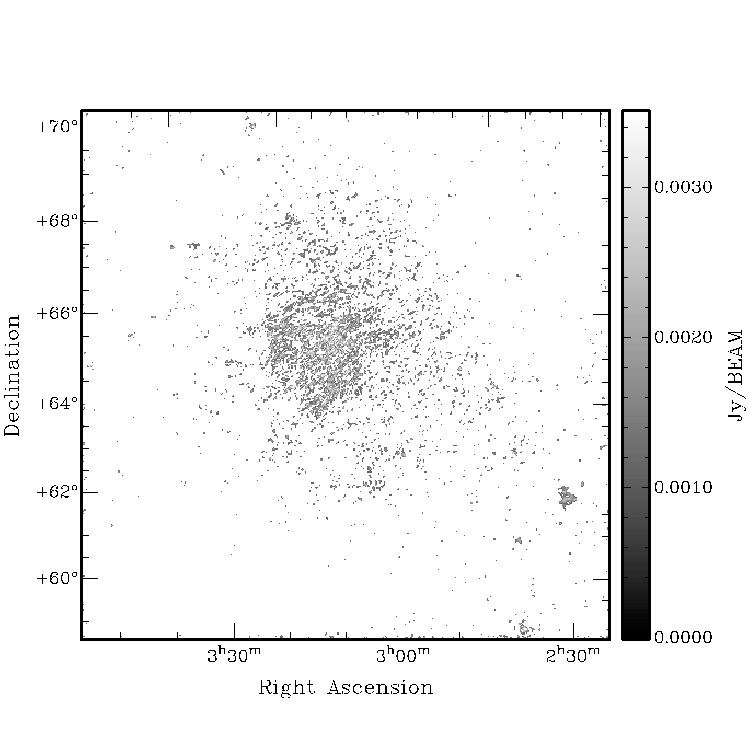}}
\resizebox{0.9\columnwidth}{!}{\includegraphics{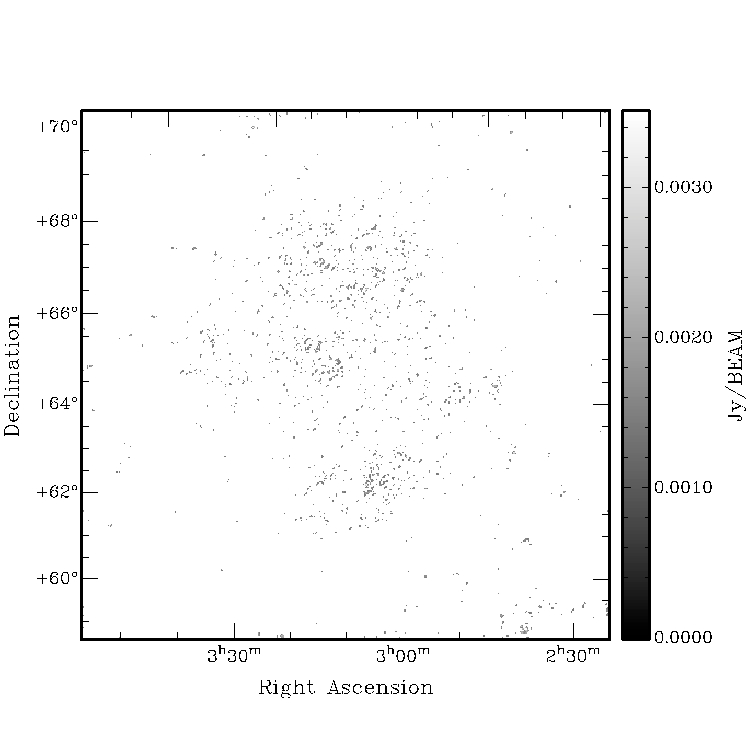}}
\resizebox{0.9\columnwidth}{!}{\includegraphics{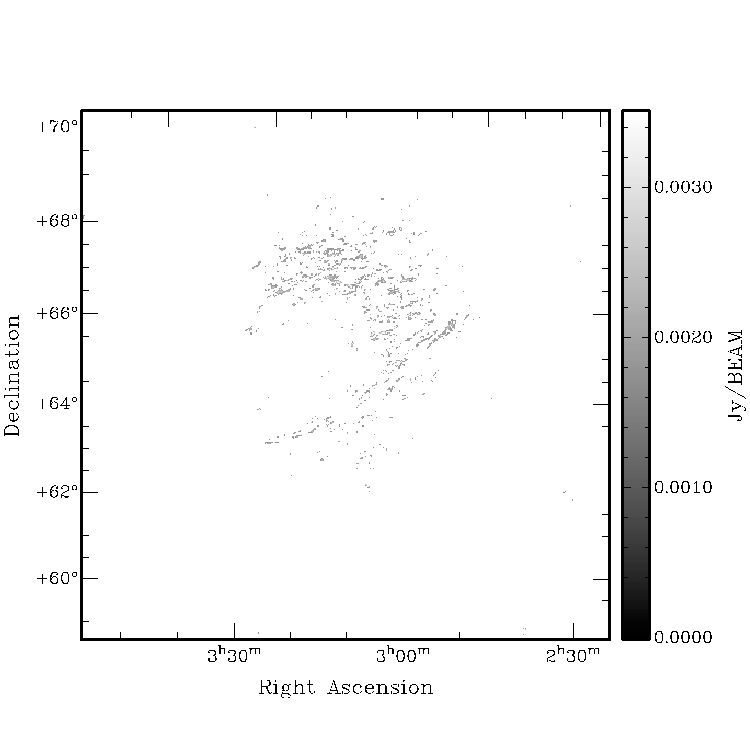}}
\caption{\label{rgb_image} \textit{Composite image showing in the upper left panel a colour coded image of polarized intensity emission clipped at $3\sigma$ at three main Faraday depth ranges, where the white crosses denote the lines of sight used to extract the Faraday dispersion spectra in figure \ref{faraday_dispersion_spectra}. PI clipped maps averaged over ranges in Faraday depth depicting the ``bubble'', the ``ring'' and the ``curtain'' are shown clockwise in grey scale; Red (upper right panel) is averaged PI over a range of $\phi \in [-13,-5]$~rad~m$^{-2}$, Green (lower right panel) is averaged PI over a range of $\phi \in [-4,-1]$~rad~m$^{-2}$ and Blue (lower left panel) is averaged PI over a range of $\phi \in [0,+5]$~rad~m$^{-2}$.}}
\end{figure*}

To emphasize properties of linearly polarized emission in $\phi$ space, the PI emission was integrated in the frames from $-13$~rad~ m$^{-2}$ to $-5$~rad~m$^{-2}$, from $-4$~rad~m$^{-2}$ to $-1$~rad~m$^{-2}$, and from $0$~rad~m$^{-2}$ to $+5$~rad~m$^{-2}$ as \citep{Brentjens07} \\
\begin{equation} P(\Delta \phi) = B^{-1} \sum_{\phi \, min}^{\phi \, max} (PI(\phi) - \sigma_{PI} \sqrt{\frac{\pi}{2}}) \: ,\end{equation} 
where the normalization factor $B$ is the area of the RMSF beam after RM-CLEAN divided by the interval between two frames of the data cube, and $\sigma_{PI}$ is the noise in PI estimated from figure~\ref{noise_profile}. These maps were clipped at a level of $3\sigma$ and then combined into a composite false-colour image as shown in the left-hand upper plot in figure~\ref{rgb_image}. \\

\begin{table*}
\label{table:2}
\caption[]{Properties of the three components.}
\centering
\begin{tabular}{lccccc}
\hline\hline
     Structure         & Colour & $\phi$~range & $\phi_{peak}$  & $<P>$ \\
                       &      & [rad~m$^{-2}$] & [rad~m$^{-2}$] & [Jy beam$^{-1}$] \\
\hline
Bubble       &$ Red $&$ 8.0 $&$ -5  $&$ (1.72\pm0.58)\times10^{-3} $ \\
Ring         &$ Green $&$ 4.0 $&$ -2  $&$ (1.78\pm0.26)\times10^{-3} $ \\
Curtain      &$ Blue $&$ 5.0 $&$ +2 $&$ (1.53\pm0.22)\times10^{-3} $ \\
\hline
\end{tabular}
\tablefoot{The columns denote: 1) Name of component, 2) Colour in figure~\ref{rgb_image}, 3) Range in Faraday depth in which component is visible, 4) Faraday depth of peak, 5) Polarized surface brightness averaged over Faraday depth range.}
\end{table*}

\begin{figure*}
\centering 
\resizebox{1.0\hsize}{!}{\includegraphics[height=10cm,width=17cm]{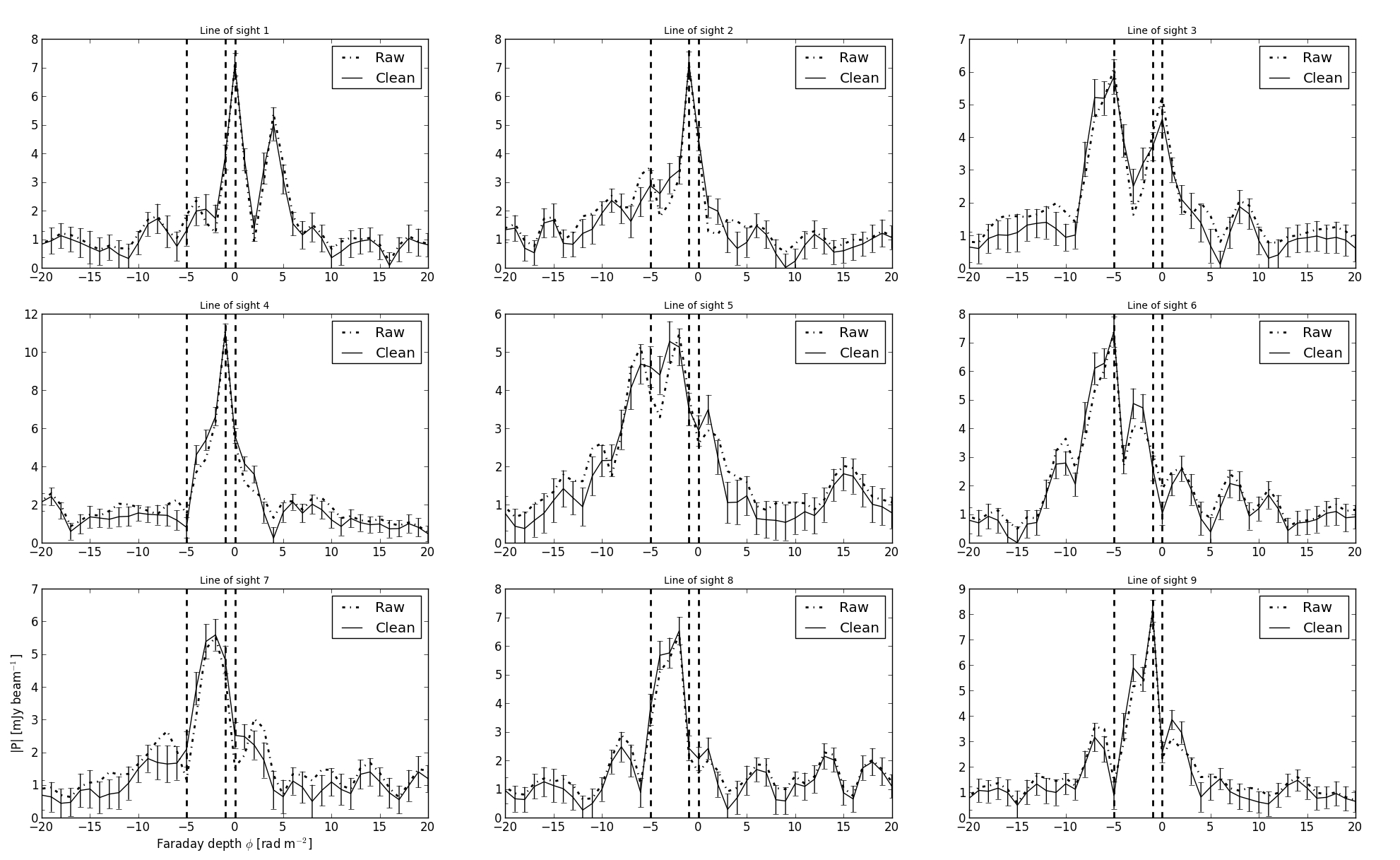}}
\caption{\label{faraday_dispersion_spectra} \textit{Faraday dispersion spectra of selected lines of sight labelled from 1 to 9 as in figure~\ref{rgb_image}. The three dashed black lines point out the position of the histogram peak for the ``bubble'', the ``ring'' and the ``curtain'' components. For each spectrum the ``Clean'' line is obtained from the ``Raw'' one by applying an RMCLEAN session with a threshold of $3$~mJy~beam$^{-1}$. The noise level is shown by $1\sigma$ error bars.}}
\end{figure*}

Figure \ref{faraday_dispersion_spectra} shows a sample of Faraday depth spectra, obtained averaging over a beam-sized box region, for several lines of sight. The lines of sight were selected to show widely varying spectra, with a strong main peak (e.g. spectrum 7) or with double and multiple peaks (e.g. spectra 1, 3, and 5 respectively). 
A peak in the Faraday depth spectrum indicates an emission component in Faraday space. However, due to the limited frequency coverage, components in Faraday space may be represented in the Faraday depth spectrum incompletely or not at all.

We estimated the mean polarized surface brightness  for each component (see~Table~2). These mean values are of the same order of magnitude. Interestingly the ``bubble'' and ``ring'' components have comparable polarized surface brightness, slightly higher than the ``curtain''. The mean polarized brightness temperature over the range $\phi \in [-13,+5]$ is $T_{b} \sim 5.6$~K.

\subsection{Polarized extragalactic background sources}
\label{s:discrete}

\begin{figure}
\resizebox{1.0\columnwidth}{!}{\includegraphics{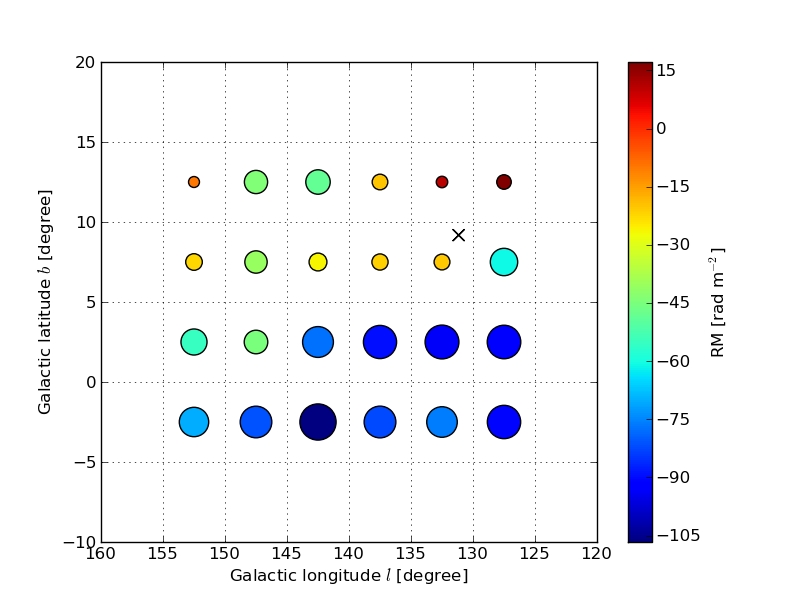}}
\caption{\label{NVSS_RM_5deg_bin} \textit{Distribution of polarized extragalactic point sources and rotation measure values from the NVSS catalogue after binning over 5~degrees spatial scale across the imaged field and its surroundings. The position of pulsar B0226+70, which is used in section 7 to constrain the spatial distance of observed polarized structures is indicated by a cross.}}
\end{figure}

Although we are primarily interested in the Galactic emission we briefly point out a few interesting discrete extragalactic objects within 5$^{\circ}$ of the field centre. These are 
\begin{itemize}
\item The spiral galaxy IC\,342, which is visible both in polarized and total intensity. Large scale disk/halo emission at $\phi = (-3.0 \pm 0.5)$~rad~m$^{-2}$ is observed, in agreement with \citet{Krause89} who find RM values between $-10$~rad~m$^{-2}$ and $10$~rad~m$^{-2}$.
\item The giant double-lobe radio galaxy WNB\,0313+683, which shows prominent linearly polarized emission from the south lobe (at $(\alpha,\delta)_{J2000} \approx (03^{h}17^{m},68^{\circ}.2)$) and the core (at $(\alpha,\delta)_{J2000} \approx (03^{h}18^{m},68^{\circ}.3)$) in a range $\phi \approx [-17,-5]$~rad~m$^{-2}$ consistent with values reported by \citet{Schoenmakers98} and \citet{Haverkorn03}. However, no polarized emission is detected from the north lobe, which is pointing away from the observer. Since the Stokes I map shows a sharp difference in brightness for the lobes (the north to south lobe mean brightness ratio is $\sim 0.2$), the non-detection of polarized emission from the north lobe can be explained as a consequence of relativistic beaming and/or additional Faraday depolarization \citep{Heald09}.
\end{itemize}
We used discrete polarized background sources to search for a latitude and/or longitude dependence of the Faraday depth across the field of view. From the NVSS RM catalogue \citep{Taylor09} we selected a sample of polarized sources with known RM, which covers the observed field of view and its surroundings (see Fig.~\ref{NVSS_RM_5deg_bin}). To average out intrinsic RM components of the extragalactic sources, we re-sampled the data over square bins with an angular scale of 5$^{\circ}$ (about half the size of the imaged field). We find a clear gradient in Galactic latitude of $\Delta RM / \Delta b \approx 4.6$~rad~m$^{-2}$~deg$^{-1}$, while no evident Galactic longitude dependence is found. The dominant negative RM values imply overall negative $B_{\parallel}$ values, in accordance with the observed negative Faraday depths. A systematic increase in RM values for sources below $b = 5^{\circ}$ is observed as expected because of the presence of the Galactic disk. In the longitude range $l \in [125^{\circ},140^{\circ}]$ the presence of extended \ion{H}{ii} regions (i.e. the W3/W4/W5/HB3 \ion{H}{ii} region/supernova remnant (SNR) complex in the Perseus arm), as well as the H$\alpha$ maps \citep{Haffner03} suggests that the increase in RM towards lower latitudes is at least partially a consequence of enhanced thermal electron density. We note that both the diffuse emission and polarized point sources show predominantly  negative Faraday depth and RM, respectively. However, RMs from the extragalactic point sources are much more negative than the Faraday depths of the diffuse emission, suggesting that (a) the diffuse emission is nearby and does not span the full line of sight through the Galaxy, and (b) that $\vert$RM$\vert$ increases through the outer Galaxy without large-scale magnetic field reversals.

\section{Model for the diffuse polarized emission}

In this section, we discuss plausible sources in the ISM that correspond to the three polarized components identified in Section~5. Interpretation of these polarized components in terms of synchrotron emitting and/or Faraday rotating regions is not straightforward for a number of reasons. First, these are interferometric data, which are missing short spacings in the uv-plane and are therefore insensitive to structure on scales larger than two degrees. This explains why a large amount of polarized emission is observed, but only very weak total intensity \citep{Bernardi09}, providing an apparent polarization degree far above 100\%. This indicates that the synchrotron radiation is emitted on large scales, but its polarization is altered on smaller scales by Faraday effects. In this case we can calculate the missing large-scale structure contribution for Stokes Q and U for a uniformly polarized background propagating through a small-scale Faraday screen \citep{Haverkorn02}. The expected offset for Stokes Q and U ($Q_{0}$ and $U_{0}$) depends on the wavelength ($\lambda$) and the width of the Faraday depth distribution ($\sigma_{\phi}$) as
\begin{equation}e^{-2\sigma_{\phi}^{2}\lambda^{4}} \rightarrow 0 \; \Rightarrow \; Q_{0}=U_{0}=0 \,.\end{equation}
At low frequency this condition is typically well satisfied and we do not expect large-scale undetected Stokes Q and U structures. Although there is missing large-scale structure in total intensity Stokes~I, there is no missing large-scale structure in Stokes~Q and U, therefore the polarization data are reliable. Second, the limited frequency range of the data set precludes the detection of Faraday depth components that have a width $\delta\phi > 1$~rad~m$^{-2}$ (Section \ref{s:rmsynth}). Therefore, by definition, the Faraday depth components that we observe are unresolved in Faraday depth. Analogously to the missing short-spacing problem, this indicates that there likely is Faraday-thick polarized emission (e.g. emission over a wide range of Faraday depths), which we cannot detect. Last, depolarization effects can decrease the polarized intensity of the observed structures.

First, we calculate some properties of the global interstellar medium in the direction of the FAN region, below. Then we discuss what information is available on the absolute and relative distances of the three components in Section~\ref{s:distance}. Finally, we present our model in Section~\ref{s:model}.

From the mean dust-corrected H$\alpha$ surface brightness across the (half upper part of the) field of view ($I_{H\alpha} \approx 3.3$~R) and adopting an electron temperature in the diffuse ionized interstellar medium $T = 8000$~K, we derive the related emission measure $EM_{field}=7.43$~pc~cm$^{-6}$. This corresponds to a large-scale mean free electron density of $n_{e} \approx 3 \times 10^{-2}$~cm$^{-3}$ over an assumed total path length of 7~kpc, a value, which is in good agreement with the literature \citep{Hill08,Ferriere01,deAvillez12}. 

The Fan region has a high degree of polarization, which indicates a relatively regular, large-scale  magnetic field topology. This can be either uniform or anisotropically random magnetic fields.

The uniform magnetic field component following the spiral arms is almost perpendicular to the line of sight in this direction. Therefore, the impact of the large-scale interstellar magnetic field on Faraday depth values toward the Fan region is expected to be small. We can test this assumption by estimating the path-averaged magnetic field amplitude along the line of sight up to a few hundred parsecs. We consider a sample of nearby pulsars around the imaged field (see Table~\ref{tabfeat}) and apply the general relation: \begin{equation}\langle B_{\parallel} \rangle = 1.24\frac{RM}{DM} \; .\end{equation} 

\begin{table*}
\caption[]{Parameters of nearby pulsars located in the direction of the FAN region.}
\centering
\label{tabfeat}
\begin{tabular}{@{}lrrrrrrrr}
\hline\hline
Object & Name & l,b & DM & RM & Dist\tablefootmark{a} & Dist DM\tablefootmark{a} & $\langle B_{\parallel} \rangle$ \\
 &    & [deg,deg] & [cm$^{-3}$ pc] & [rad m$^{-2}$] & [kpc] & [kpc] &  [$\mu$G] \\
\hline
1 & B0450+55 & 152.62\,,\,7.55 & $14.495\pm7$ & $10.0\pm3$ & $1.19$ & $0.79$ & $+0.8555$ \\
2 & B0655+64 & 151.55\,,\,25.24& $8.771\pm5$ & $-7.0\pm6$ & $0.48$ & $0.48$ & $-0.9896$ \\
3 & B0809+74 & 140.00\,,\,31.62& $6.116\pm18$ & $-11.7\pm13$ & $0.43$ & $0.33$ & $-2.3721$ \\
4 & B1112+50 & 154.41\,,\,60.36& $9.195\pm8$ & $3.2\pm5$ & $0.54$ & $0.54$ & $+0.4315$ \\
\hline
\end{tabular}
\tablefoot{For each object the inferred magnetic field component along the line of sight $\langle B_{\parallel} \rangle$ is also listed. Data are taken from the ATNF Pulsar Database \citep{Manchester05}.\\
\tablefoottext{a}{Dist is the best estimate to the pulsar distance (kpc).}
\tablefoottext{b}{DistDM is the distance based on the \citet{Taylor93} electron density model.}}
\end{table*}

With the exception of object 1, all pulsars have estimated distances lower than $\approx 600$~pc. The scatter in $B_{\parallel}$ is high, and its sign changes for different pulsars, confirming that the uniform magnetic field component is much smaller than the random component in this direction. We calculate a mean parallel magnetic field $\langle B_{\parallel} \rangle \approx 1.2$~$\mu$G.

\subsection{Distance estimates to the structures}
\label{s:distance}

Most of the polarized emission can be found at slightly negative Faraday depths, consistent with earlier observations \citep{BrouwSpoelstra76,Taylor09}, which suggests that the radiation is emitted at a short distance from the observer. \citet{Wilkinson74} and \citet{Uyaniker03} deduce that this polarized emission and the low RM values in the second quadrant around $l \sim 150^{\circ}$ must be nearby, possibly at a distance $\lesssim 500$~pc.  \\ 
Also, the NVSS gradient show a mean RM value at the FAN latitude of $\overline{RM}_{b=7^{\circ}} \approx -32$~rad~m$^{-2}$, which is higher than the observed Faraday depths. We conclude that the detected structures are not representative of the general behaviour through the entire line of sight; as a consequence, these structures must be close.

We can roughly estimate the distance to the  ``bubble'', assuming uniform conditions for the medium along the entire line of sight, as
\begin{equation} d_{D} = \frac{\phi_{OBS}}{\phi_{REF}} D_{REF}[pc] \: ,\end{equation} where $\phi_{OBS}$ is the measured Faraday depth, and $\phi_{REF}$ the reference value of Faraday depth (or rotation measure) corresponding to the line of sight total length $D_{REF}$.
We consider the rotation measure of nearby radio pulsar B0226+70 at $(\alpha,\delta)_{J2000} \approx (02^{h}31^{m},70^{\circ}.26)$ as a reference. The distance of this pulsar is $d_{PSR} = (2.25 \pm 0.56)$~kpcn and its rotation measure $RM_{PSR} = (-56 \pm 21)$~rad~m$^{-2}$ from the ATNF Pulsar Database\footnote{http://www.atnf.csiro.au/research/pulsar/psrcat/} \citep{Manchester05}. Equation (8) then gives $d_{BUBBLE} \approx (201 \pm 90)$~pc, which is consistent with a previous estimate \citep{Haverkorn03}. At such a distance and considering the angular size of the ``bubble'' ($\approx 3^{\circ}$), a physical size in the plane of the sky of $\approx 10$~pc is obtained. 

Considering the intrinsic Faraday depth for the ``curtain'' and ``ring'' components along with the large-scale electron density field and magnetic field component estimated above along the line of sight, we obtain an estimate of the distance needed to build up this Faraday depth. We find ranges of $\approx 70$~pc and $\approx 140$~pc for the ``curtain'' and the ``ring'', respectively. These distance estimates do have considerable error margins due to the strong assumptions.

These distance estimates can be tested by comparing the amount of polarized intensity built up over that distance to the observed polarized intensity in the component. At 408~MHz, the total intensity emissivity is $\epsilon_{I}^{408}\sim(11\pm3)$~K~kpc$^{-1}$ or 0.011~K~pc$^{-1}$ \citep{Beuermann85}. With a spectral index of $-2.5$, the conversion factor from 408 to 150~MHz is $\sim12.2$. Thus the total intensity emissivity at 150~MHz is $\epsilon_{I}^{150}\sim12.2\times\epsilon_{I}^{408}=(134\pm37)$~K~kpc$^{-1}$ or 0.13~K~pc$^{-1}$. If we assume a polarization degree of $\sim8\%$ \citep{Haverkorn08}, then $\epsilon_{PI}^{150}\sim\epsilon_{I}^{150}\times0.08=(10.72\pm2.96)$~K~kpc$^{-1}$ or 0.0094~K~pc$^{-1}$. Therefore, for a constant $\epsilon_{PI}^{150}$ (which implies a constant polarization degree of 7\%) over the line of sight through the ``curtain'' component, we need about $(140\pm39)$~pc to build up the polarized intensity of the ``curtain'' of $\sim 1.2$~K. This estimate is higher than our estimate above, probably because the ``curtain'' component is likely to show enhanced synchrotron emission in a narrow emission region (see Section~\ref{s:curtain}). The ``ring'' component would need about $(168\pm46)$~pc to build up the proper polarized intensity, consistent with the estimate above.

We can also give information about the relative positions and distances of the components.  The ``curtain'' component is the nearest component. Since it does not show any imprint of the ``bubble'', it has to be located in front of it. The observation of small Faraday depths in this component suggests that it is the nearest layer of ISM with respect to the observer.

The Faraday depths corresponding to these three components are in adjacent ranges, suggesting that their relative distances are small. In particular, the ``bubble'' has to be located close behind the ``curtain'' . This is because an additional synchrotron emission component would build up in a large distance between ``bubble'' and ``curtain'', which would have been observable. The morphology of PI patterns, as well as the very similar mean polarized surface brightness (see Table~2), point to a clear link between the ``bubble'' and the ``ring''. 

\subsection{The foreground ``curtain'' component}
\label{s:curtain}

The change in sign of Faraday depth between the ``curtain'' and the other components indicates a reversal of the magnetic field component along the line of sight $B_{\parallel}$ behind the ``curtain''.  A plausible magnetic field configuration of how this can be achieved is, e.g., a localized, ordered, magnetic field in the Galactic plane perpendicular to the observer, which fans out locally to produce oppositely directed parallel field components. This magnetic field configuration has a maximum synchrotron emission at the location of the perpendicular magnetic field lines, possibly corresponding to the synchrotron emission from the ``curtain''.

\begin{figure}
\resizebox{1.0\columnwidth}{!}{\includegraphics{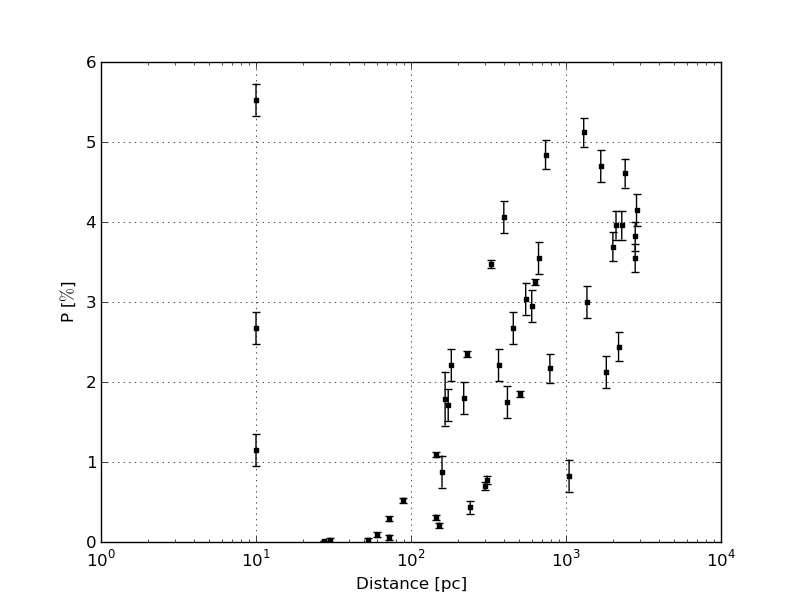}}
\caption{\label{opt_pol} \textit{Increase in the polarization degree of optical starlight from stars within the Fan field, closer than 3~kpc.}}
\end{figure}

We propose that this component corresponds to the wall of the Local Bubble. The above size estimates for this foreground component roughly agree with the dimension of the Local Bubble in this direction \citep{Snowden98}. A tomography of the interstellar gas within 250~pc was recently obtained by \citet{Welsh10} combining interstellar absorption data for stars with a Hipparcos measured parallax. Maps of spatial distribution for the neutral gas and neutral plus ionized gas components were derived from \ion{Na}{i}, \ion{Ca}{ii} tracers, displaying the Local Bubble as a region that is poor in neutral gas (and dust) but with several diffuse and highly ionized clouds. A visual inspection of both \ion{Na}{i} and \ion{Ca}{ii} galactic projected maps (see their Figs. 12 and 15) toward the direction $l = 137^{\circ}$, $b = +7^{\circ}$ reveals the presence of an extended and high-density wall of partially ionized medium at about 100 pc of the Sun, as well as a rarefied medium inside the wall, in agreement with our estimate. Moreover, optical starlight measurements in the Fan region from \citet{Heiles00} show a rise in polarization degree from a distance $\sim 100$~pc, as shown in Fig.~\ref{opt_pol}. This can be explained by an increase in the dust density, denoting the edge of the Local Bubble. The few high polarization degrees at $\sim 10$~pc can be explained by a compact and cold cloud in the very local interstellar medium. In this picture, the magnetic field would be aligned with the Local Bubble wall, i.e.\ mostly perpendicular to the line of sight, which indicates enhanced synchrotron emission in the Bubble wall. The Faraday rotation of that emission by the Local Bubble interior is small, i.e. $\lesssim 2$~rad~m$^{-2}$. Using the computed mean parallel magnetic field component of $\langle B_{\parallel} \rangle \approx 1.2$~$\mu$G and a path length of 70~pc, the mean electron density is $n_{e}^{LB} \approx 1.5 \times 10^{-2}$~cm$^{-3}$ in the Local Bubble, which agrees with literature \citep{Spangler09}.

Previously, an association of the extended polarized Galactic radio synchrotron foreground with the Local Bubble was made by \citet{Brentjens11}, who observed the area around the Perseus cluster near $l = 150^{\circ}$, $b = -13^{\circ}$ at $\lambda = 0.80-0.88$~m. In this region, negative Faraday depth values are expected from the inferred direction of the large-scale magnetic field, while he finds a positive Faraday depth component at $\phi$ = +6~rad~m$^{-2}$ and estimates its size of approximately 200~pc. These values differ from our estimations by about a factor two and may indicate density variations and irregular shape of the Local Bubble wall.

\subsection{The background ``ring'' component}
\label{s:background}

The ``ring'' component shows a small south-east to north-west gradient in Faraday depth ($\lesssim 1$~rad~m$^{-2}$~degree$^{-1}$). This  gradient does not correspond to the large-scale gradient in RM expected from the NVSS data \citep{Taylor09}, but is directed almost perpendicular to that. If, as we argue below, the ``bubble'' component is just an additional Faraday rotation of synchrotron emission in the ``ring'' component, we would expect this gradient across the field to also be present in the ``bubble''. However, since the Faraday depth gradient of $\sim 2$~rad~m$^{-2}$ is measured over the entire imaged field, the expected $\phi$ gradient over the bubble size (about 1/3 of the entire field) turns out to be less than $1$~rad~m$^{-2}$~degree$^{-1}$ (i.e. less than 1/3 of RMSF width), hence undetectable with our observational set up.

\subsection{The ``bubble'' component}

The ``bubble'' component is detected in polarized intensity, but not in total intensity. This could be due to the lack of short spacings. However, if the bubble were significantly emitting, its polarized intensity would be higher than the ring component. Instead, their polarized brightness temperatures are comparable, suggesting that the ``bubble'' is a Faraday rotating structure, retro-illuminated by the synchrotron emission from the ``ring''. 

We interpret these observations as a consequence of the configuration with the ``bubble'' located in the spatially extended ``ring'' structure. Two configurations are possible for the ``bubble'': it can be assumed to be either a filled sphere or a shell surrounding a density depletion. We prefer the first configuration because PI emission at the greatest Faraday depths is observed in the centre of the ``bubble''. Indeed under the assumption of pressure balance a depletion of warm ionized medium relative to its surroundings may imply an enhancement of neutral gas or dust, which is not observed (e.g. the total \ion{H}{i} or IRAS 100~$\mu$m maps). Instead since the early work of \citet{Verschuur69}, a considerable lack of \ion{H}{i} is observed, as well a very good agreement between higher RM / Faraday depth values and lack of \ion{H}{i}.  Also, for a filled sphere it is easier to build up the higher Faraday depths observed through the ``bubble''

The additional Faraday rotation in the ``bubble'' component can be caused by increased $B_{\parallel}$, increased $n_e$, or a combination of these two options. Observations of H$\alpha$ emission measures set an upper limit for the increase in $n_e$ in the bubble. The bubble is not detected \citep{Haverkorn03} in the WHAM Northern Sky Survey \citep{Haffner03}, which has a sensitivity of 0.05~R. For the emission not to be detected at $3\sigma$ (i.e. 0.15~R), the emission measure has to be $EM_{3\sigma} \lesssim 0.34$~pc~cm$^{-6}$. 

\begin{figure}
\resizebox{1.0\columnwidth}{!}{\includegraphics{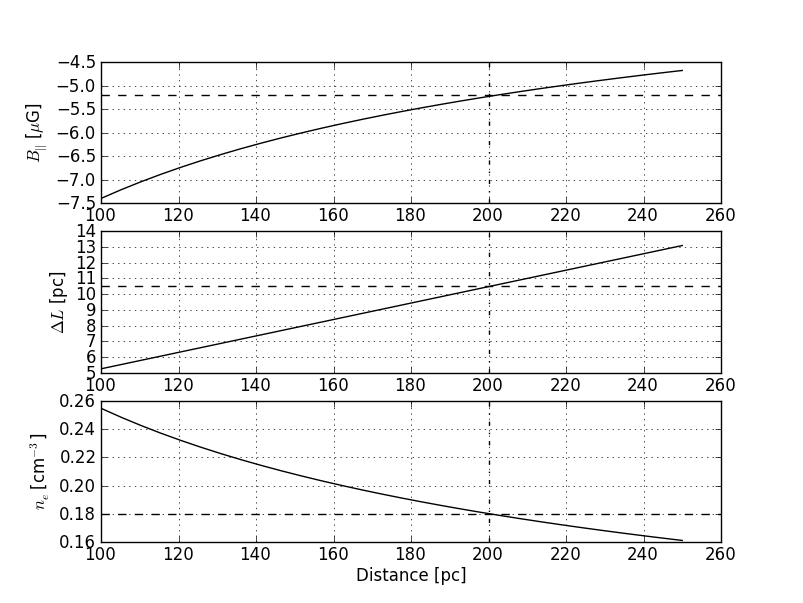}}
\caption{\label{local_warm_cloud} \textit{Variation in $B_{\parallel,Bubble}$, $n_{e,Bubble}$, and $\Delta L_{Bubble}$ as a function of ``bubble'' distance. Also, for a reference distance of $\sim200$~pc (vertical dot dashed black lines) the upper limit for the ``bubble'' electron density ($n_{e,Bubble}\sim0.18$~cm$^{-3}$), the magnetic field component along the line of sight $B_{\parallel,Bubble}\sim5.25$~$\mu$G and its size $\Delta L_{Bubble} \sim 10.5$~pc are indicated by the horizontal black dashed lines.}}
\end{figure}

Assuming that the bubble is spherical, the magnetic field component along the line of sight $B_{\parallel}$ can be estimated as \begin{equation} \Delta\phi_{max,Bubble} \approx 0.81 \times n_{e,Bubble} \times B_{\parallel,Bubble} \times \Delta L_{Bubble}\:. \end{equation} We can express both the size and the free electron density as a function of the distance to the ``bubble'': $\Delta L_{Bubble} \simeq \tan 3^{\circ} \times d_{Bubble}$ and $n_{e,Bubble} = \sqrt{EM/\Delta L_{Bubble}}$. By using these expressions we can check the variation in the magnetic field component along the line of sight, the free electron density and the size of the ``bubble'' as a function of its distance, indeed indicating an enhancement (with respect to the mean values computed above) of both the free electron density and magnetic field strength along the line of sight by a factor of a few as shown in figure \ref{local_warm_cloud}. A reference distance of about 200~pc implies a path length through the ``bubble'' of $\lesssim 11$~pc and the corresponding electron density and magnetic field strength within the bubble feature are $n_{e,Bubble}\sim0.18$~cm$^{-3}$ and $B_{\parallel,Bubble}\sim5.25$~$\mu$G, respectively.

\begin{figure*}
\centering
\resizebox{0.9\columnwidth}{!}{\includegraphics{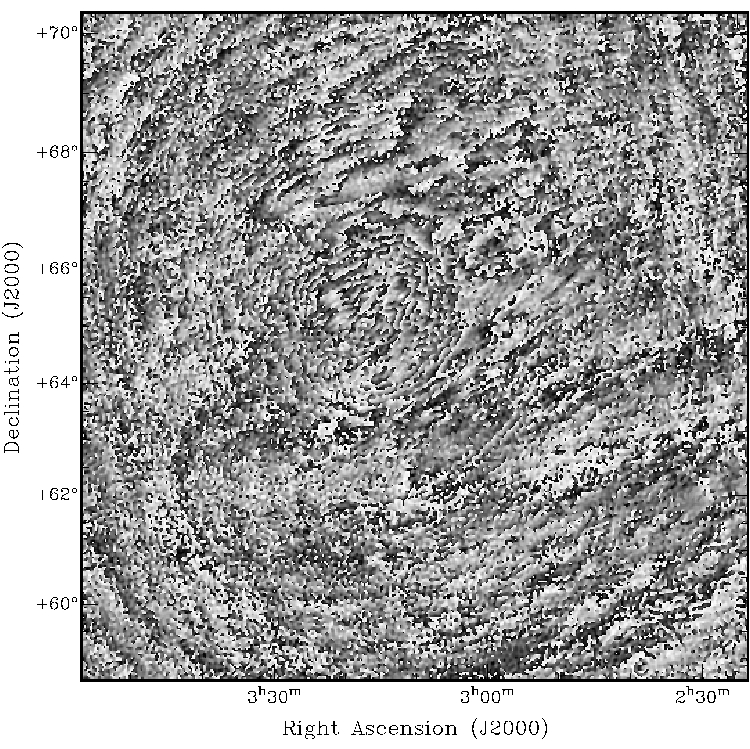}}
\resizebox{0.9\columnwidth}{!}{\includegraphics{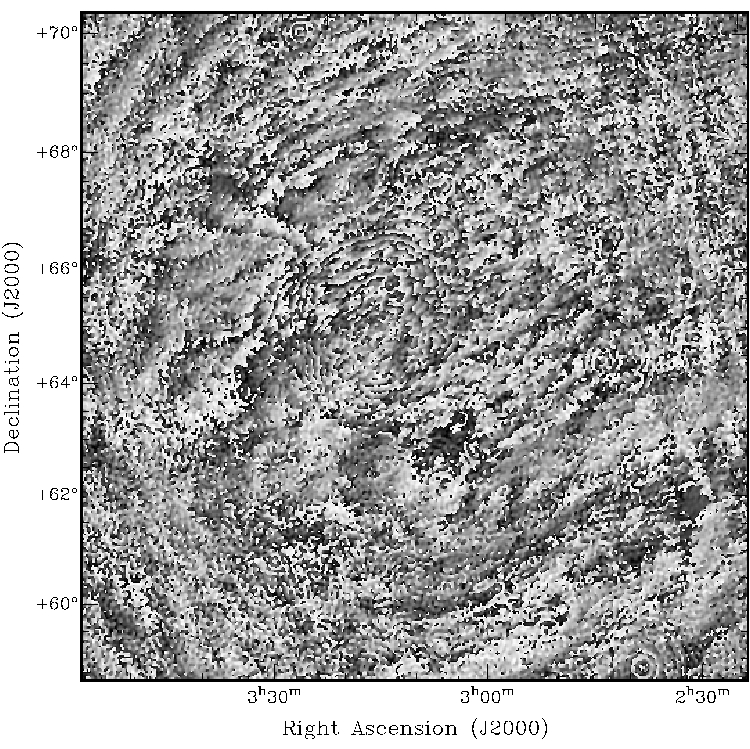}}
\resizebox{0.9\columnwidth}{!}{\includegraphics{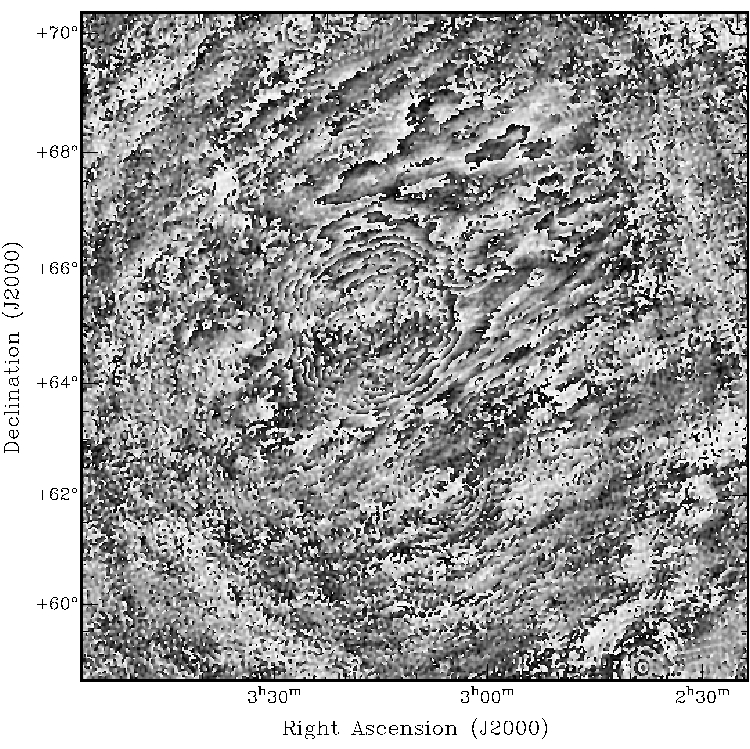}}
\resizebox{0.9\columnwidth}{!}{\includegraphics{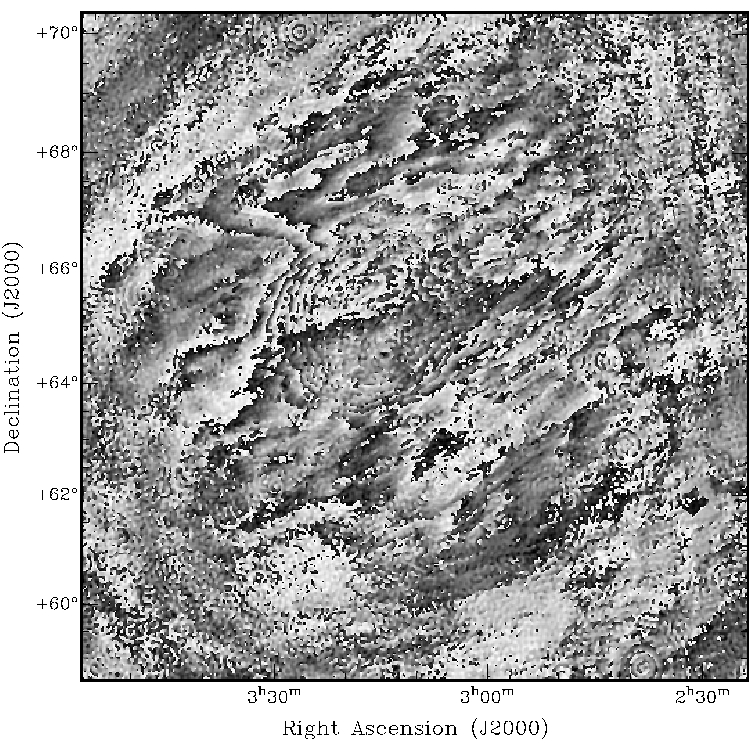}}
\caption{\label{pol_angle} \textit{Grey scale images of polarization angle at selected Faraday depths at $\phi = [-7,-5,-3,-1]$~rad~m$^{-2}$ from top left to bottom right. The gray scale runs from +90$^{\circ}$ (black) to -90$^{\circ}$ (white).}}
\end{figure*}

Extra morphology information is obtained from the polarization angle maps. A highly structured configuration for the projected field is observed in the $\phi \in [-10,0]$~rad~m$^{-2}$ range. As can be seen from figure \ref{pol_angle}, the polarization angles (also in Stokes Q,U) show very ordered structures around the edge of the bubble. The circular, rapid variations in angle are due to a large angle gradient, displayed in a fixed angle range of $[-0.5\pi, 0.5\pi]$~rad. Each "ripple" corresponds to an angle gradient over $\sim 8$~arcmin, followed by an abrupt jump across $\sim 2$~arcmin, which is due to the $n\pi$-ambiguity. The polarization angle thus varies by $\pi$ radians over $\sim10$~arcmin. On average four or five of these ripples are visible, which indicates an angle gradient of $\Delta\theta \sim 5\pi$~rad over 50~arcmin. At a wavelength of two metres, this indicates a gradient in Faraday depth $\Delta\phi = \Delta\theta / \lambda^2 \approx 3.9$~rad~m$^{-2}$ over 50~arcmin, corresponding to $\approx2$~pc at the estimated distance of the ``bubble''.

Moreover, we can use the Faraday depth gradient derived from the rippled features in the polarization angle map to estimate the maximum $\phi$ value through the ``bubble''. Assuming uniform conditions within the ``bubble'' for the magnetic field and the thermal electron density, we find the total amount of Faraday depth to be $\sim 5$~rad~m$^{-2}$, which agrees with the observed value. Finally a lower Faraday depth value for the nearby surrounding ``Fan'' medium is well supported by previous observations \citep{BrouwSpoelstra76}.

\begin{figure}
\resizebox{1.0\columnwidth}{!}{\includegraphics{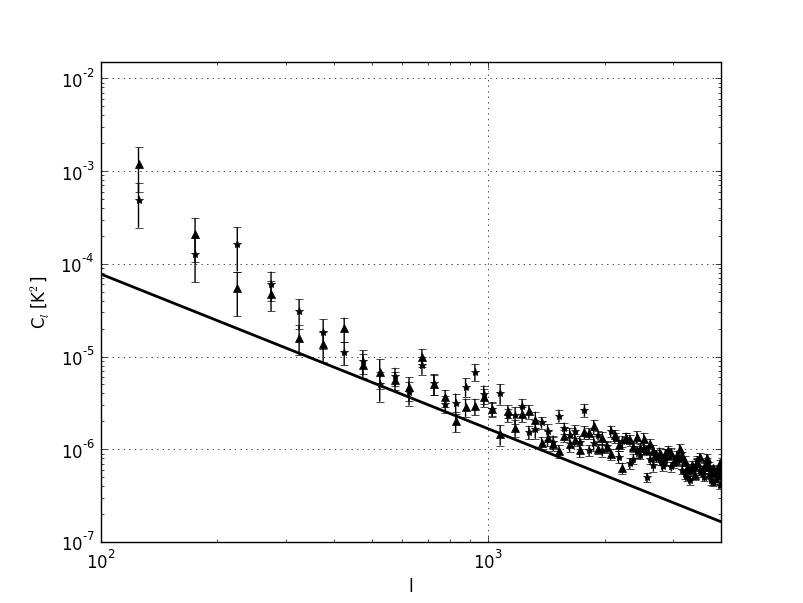}}
\caption{\label{pi_powspec} \textit{Power spectral density of polarized intensity of the ``ring'' (stars) and ``bubble'' (circles) components as derived from a 6 degree inner square of the PI images integrated over the relevant Faraday depths. Also statistical errors and the Kolmogorov spectral slope are shown by $1\sigma$ error bars and a black solid line, respectively.}}
\end{figure}

The chaotic and patchy appearance of observed features both in polarized intensity and polarization angle imply a large amount of depolarization and may be the result of the turbulent state of the probed magnetized medium. The power-law behaviour of the power spectrum indicates a turbulent medium. In figure \ref{pi_powspec} we show power spectra of the bubble and background components; the spectral energy distributions are close to, but slightly flatter than, a Kolmogorov spectrum. Furthermore, the presence of an inertial range is apparent, which extends for more than a decade\footnote{Due to the finite size of the field of $\sim 12^{\circ}$, the smallest mode we can consider is $k_{min} \sim 10^{2}$, while the largest mode $k_{max} \sim 2.6 \times 10^{3}$ is defined by the angular resolution.} and flattens at modes $k\gtrsim 10^{3}$ . Such an increase in power on small scales agrees with the expected large amount of depolarization.\\

\subsection{The model in physical space}
\label{s:model}

\begin{figure}
\resizebox{1.0\columnwidth}{!}{\includegraphics{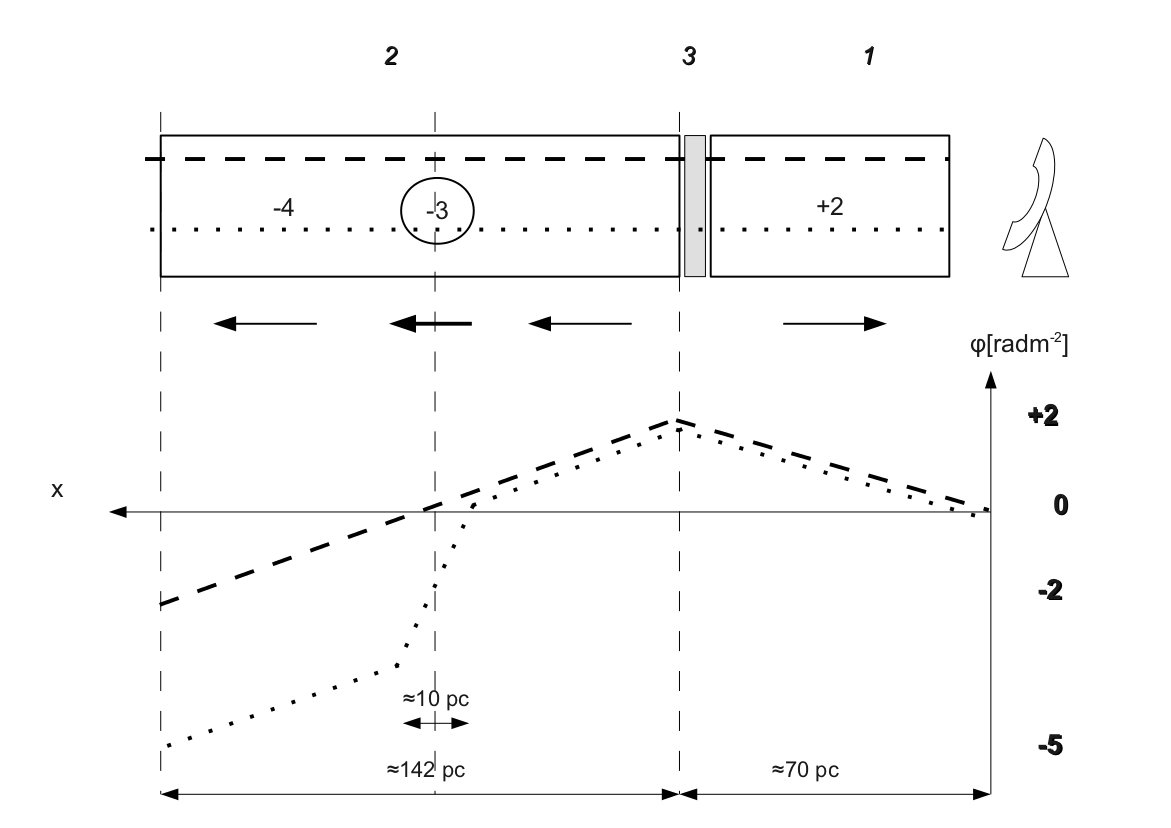}}
\caption{\label{model_fan} \textit{Cartoon illustrating the proposed model for the observed distribution of Faraday depths. Grey areas indicate only emitting regions, and white blocks indicate regions both emitting and Faraday rotating. Intrinsic Faraday depth values are shown within each structure in the top panel, and observed Faraday depth profiles are built up in the lower panel for two representative lines of sight: the dotted line crosses all the detected structures, the dashed line misses the ``bubble''. Arrows in the top panel point out the magnitude and sign of $n_{e} B_{\parallel}$ }}
\end{figure}

These considerations lead us to a qualitative model in physical space as depicted in figure~\ref{model_fan}. The region nearest to the observer is bordered by a synchrotron emission layer that corresponds to the Local Bubble wall, about 70~pc away (the ``curtain''). This region has a small Faraday depth and is therefore (partially) ionized and has a parallel magnetic field component towards the observer. The border region has enhanced synchrotron emission, which points to an increase in magnetic field perpendicular to the observer.

Behind this region is an extended, magnetized region corresponding to the ``ring'', a volume of warm partially ionized and weakly emitting medium which contains the ``bubble'' component. The parallel component of the large-scale magnetic field in this direction is expected to point slightly away from us \citep{Sun08}, in agreement with these two components at negative Faraday depths. Since most of the diffuse synchrotron emission is thought to be emitted in the spiral arms and the spatial distances we infer are within a few hundreds of parsec, we conclude that the overall detected PI emission originates primarily in the Local arm. 

\subsection{Possible associations}

\begin{table*}
\label{table:stars}
\caption[]{Main properties of selected stars within the ``bubble''. Data are taken from the SIMBAD database.}
\centering
\begin{tabular}{lcccccc}
\hline\hline
Star ID& Spectral type& $\alpha$ & $\delta$ & Distance &Prop. motions &Radial vel. \\
 o    &      		  & (J2000) & (J2000) &   [pc]   &  [mas/yr]   &    [km/s] \\
\hline
WD\,0314+64  & DA	 &03 18 35.07 & $+$65 00 01.2 &  &110.0,$-$169.0 & \\
HIP\,15520  &B2\,IV:e  &03 19 59.273 & $+$65 39 08.250 & 234$\pm$24 &11.25,$-$14.56 & $-$3.4 \\
\hline
\hline
\end{tabular}
\end{table*}


Circularly shaped, polarized objects without counterparts in other tracers have been observed before \citep[e.g.][]{Gray98,Uyaniker02}. These observations differ in two ways from the previous ones.

Firstly, we observe much more small-scale fluctuations in the polarization angle patterns. However, this can be caused by the different observing frequencies: Faraday effects are much stronger at low frequencies, so that angle variations are expected to be more evident. Secondly, \citet{Gray98} and \citet{Uyaniker02} explain their polarization structures in terms of enhancement in electron density, while we \citep[and][]{Haverkorn03} argue that some of the structure must be due to enhanced magnetic field. The importance of the magnetic effects and their coupling with the particle density is supported by the ratio of thermal to magnetic pressure $\beta_{pr}$ which is given by \begin{equation} \beta_{pr} = \frac{16 \pi n_{e} k T_{e}}{B^{2}} \end{equation} where $B \gtrsim B_{\parallel}$ is the total magnetic field strength, \textit{k} the Boltzmann constant, and $T_{e}$ the electron temperature in K. 
For the background ``ring'' component, with $T_{e}=8000$~K, $B_{\parallel,ring} \sim 1.2$~$\mu$G, and $n_{e} \sim 3\times10^{-2}$~cm$^{-3}$, we find $\beta_{pr} \lesssim 1.15$, which indicates a non-negligible role played by the magnetic field with respect the thermal motion. For the bubble (with $n_{e} \lesssim n_{e,Bubble}$ and $B_{\parallel} \lesssim B_{\parallel,Bubble}$) we find a lower value $\beta_{pr} \lesssim 3.62 \times 10^{-1} (T_{e}/8000~\mbox{K})$, indicating that magnetic effects dominate thermal effects over a wide range of temperatures (i.e. $T_{e} \lesssim 2.2\times10^{4}$~K); therefore the observed morphology can be shaped by the magnetic pressure of the surrounding ISM. Because of the low thermal electron density, the estimated $\beta_{pr}$ value is far outside the commonly observed range ($\beta_{pr} > 1$) for an \ion{H}{ii} region \citep{Harvey11} unless a strong gradient of temperature with respect to the surrounding is considered. \\
To address the origin of the bubble we now examine several possible origins of this component: an old supernova remnant, an old planetary nebula, and a photo-ionized region.

\subsubsection{Old supernova remnant}

The presence of a Faraday depth gradient across the bubble, the abundance of small scales and an evident polarized intensity pattern could indicate a supernova remnant. The absence of Stokes~I emission, as well as the low ratio of thermal to magnetic pressure could be explained by a very old remnant. However, no traces of shocks are found at other wavelengths, and the radial Faraday depth gradient within the bubble does not agree with the expected $\phi$-profile for supernova remnants \citep{Uyaniker02snr,Harvey10}. Finally, a supernova remnant is a shell, while the bubble is thought to be a filled spherical structure.

\subsubsection{\ion{H}{ii} region}

\begin{figure}
\resizebox{1.0\columnwidth}{!}{\includegraphics{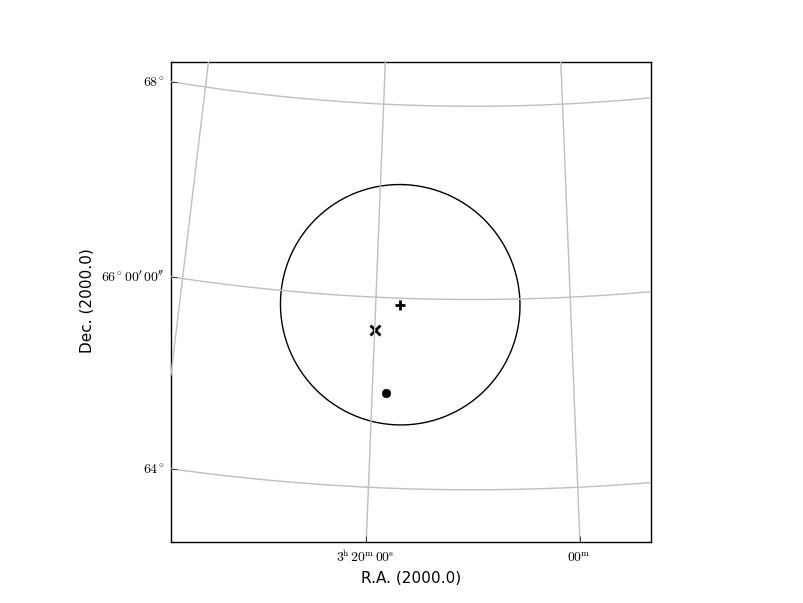}}
\caption{\label{stars_offset} \textit{Relative position of selected stars (HIP\,15520 is given by the cross, WD\,0314+64 by the dot) with respect to the polarized bubble here indicated by the circle. The offsets from the bubble centre (indicated by a plus) are $\sim 0.23^{\circ}$ and $\sim 0.83^{\circ}$, respectively, while the bubble radius is $\sim 1.15^{\circ}$.}}
\end{figure}

We consider the case of an \ion{H}{ii} region in an unusually low-density medium and a possible association with the nearby B-type star HIP\,15520 $d_{LOS} \approx (234 \pm 24)$~pc \citep{van Leeuwen07}. The distance, position and proper motion of this star (see Table~4 and figure \ref{stars_offset}) lie within the constraints for the bubble. A link between this star and the Fan region was first suggested by \citet{Verschuur69}, who pointed out the neutral gas depletion around the position of this star. He explained the disturbance in the \ion{H}{i} by the motion of this star through the interstellar medium.

However, an active \ion{H}{ii} region is a strong source of H$\alpha$. We calculated the estimates H$\alpha$ flux from the bubble if it is an \ion{H}{ii} region. The ionization flux ($Q_{0}$) for a a B2 type is $Q_{0} \lesssim 10^{45}$~ph~s$^{-1}$ \citep{Sternberg03}, which can be used to constrain the density in the Str\"{o}mgren sphere as
\begin{equation} 
n_{s} \approx 10^{3} \mbox{cm}^{-3}\left(\frac{1.2~\mbox{pc}}{R_{s}} \right)^{3/2} \times \left(\frac{Q_{0}}{5 \times 10^{49}~\mbox{photons~s}^{-1}}\right)^{1/2} \: .
\end{equation}
At a distance of 234~pc and for an angular size of $\sim 3^{\circ}$, we find a corresponding radius of 3.9~pc, a density of $\sim0.4$~cm$^{-3}$ and an emission measure of $EM_{HIP\,15520}\sim5.25$~pc~cm$^{-6}$. This expected emission measure value is  well above the WHAM sensitivity, thus it appears that a steady \ion{H}{ii} region is not likely to be responsible for the observed bubble.

\subsubsection{Old planetary nebula}

A more likely scenario would be provided by the final stage of the life of a star, forming a compact object. For a white dwarf star, some outer shells of materials have to be expelled in the surrounding ISM before the collapse by a weak shock. After this ISM sweeping stage, the hot star can ionize the surrounding medium quickly, producing a photo-ionized region. Indeed evolved planetary nebula nuclei can have Lyman continuum photon luminosities comparable to those of early B stars \citep{Reynolds86}, enabling rapid ionization of the surroundings. The observational work of \citet{Tweedy96}, focussed on old planetary nebulae, indicates there is a class of nebulae that appear to be filled, in agreement with the bubble. 

Indeed, the white dwarf star WD\,0314+648 (see Table~4 and figure \ref{stars_offset}) on the edge of the bubble, moving outwards, is an obvious first candidate. Its distance and radial motion are not known. However, the white dwarf has a high proper motion, which can be due to a small distance to the observer, or due to a kick acceleration at its formation. If we assume that this white dwarf was born at the approximate centre of the bubble, we can obtain an estimate of its age as the time needed for the star travelling from the bubble centre. We find an age of $\gtrsim 15000$~yr for the star WD\,0314+648 close to the edge of the bubble (for a proper motion and an estimated radial distance from the bubble centre of $pm^{*} \sim0.2$~arcsec~yr$^{-1}$ \citep{Lepine05} and $r^{*} \sim 0.83^{\circ}$, respectively). This age is much shorter than the recombination time, thus supporting the inferred high ionization degree. 

The effective temperature for white dwarf WD\,0314+648 is 17500~K \citep{Sion88}. This temperature is so low that the white dwarf will not ionize its surroundings, in agreement with the lack of any bubble features around WD\,0314+648. However, for it to be able to explain the bubble, it must have been much hotter and ionizing its surroundings when it was located near the centre of the current bubble.

From the above reasoning, it is possible that the bubble is a fossil Str\"{o}mgren sphere created out of an old planetary nebula and associated with a white dwarf star. Owing to the old age of the relic, interaction with the surrounding interstellar medium is expected, and the high symmetry of the observed morphology suggests that the surrounding medium is rather uniform.
For a white dwarf at a distance of 200~pc, with a density of the bubble of $n_e \lesssim 0.18$~cm$^{-3}$, the required ionization flux $Q_{0}$ is $Q_{0} \lesssim 10^{44}$~ph~s$^{-1}$. This corresponds to an emission measure $EM \lesssim 0.34$~pc~cm$^{-6}$, consistent with observations. 

From the Lyman continuum fluxes of \citet{Panagia73} and the white dwarf temperature and radius, we estimate the minimum temperature needed for sustaining the bubble: $T_{min}^{WD} \gtrsim 26000$~K. The time needed for the white dwarf to cool from $T_{min}^{WD}$ to the current value is $\Delta t \sim 10^{7}$~yr (roughly estimated from the cooling time in \citet{Mestel52}), which does not agree with the age estimated from its proper motion. Therefore, this particular white dwarf is probably not the cause of the bubble. However, other white dwarfs, at even greater angular distances from the centre of the bubble, may have created the bubble.

\subsection{Comparison of the model at other wavelengths}
We compare our model to two radio observations of this field at higher frequencies, the 85~cm WSRT observation by \citet{Haverkorn03} and the PI survey at 21~cm by \citet{Wolleben06}. In our framework the bulk of the emission must build up within and beyond the foreground ``ring'' component, at a few hundred parsec from the observer, in agreement with the conclusion of \citet{Wolleben06}. Also, in agreement with their claim of multiple emitting components over a wide range of spatial scales,  we suggest there are at least two emitting and Faraday thin components: the very nearby ``curtain'' component and the ``ring'' component; however, the ``bubble'' component is not seen in their polarized intensity map. We explain this lack of detection as due to the low synchrotron emissivity of the ``bubble'' feature at this frequency (the ``bubble'' does not manifest in emission) and to the low depolarizing impact it has on the background emission. For a maximum $\phi_{bubble}\sim8$~rad~m$^{-2}$ the expected modulation of the polarization angle at 21~cm is $\sim21$~degrees, but over the beam of the map the mean $\phi_{bubble}$ is $\lesssim4$~rad~m$^{-2}$ providing a variation in the polarization angle $\lesssim10$~degrees. Interesting is also the comparison with the 85~cm data of \citet{Haverkorn03}. Indeed, the standard RM values estimated by \citet{Haverkorn03} are in good agreement with the Faraday depths of the ``bubble'' and ``ring'' components we present in this paper. Moreover, the spatial distribution of RM values matches the observed Faraday depths observed at 2~m. However, the PI map at 85~cm displays a ring-like feature instead of a filled one. This difference is not likely to be explained by beam or depth depolarization (the resolution of PI maps is similar at 1~m and 2~m) but may be the consequence of the missing short spacing on Stokes Q and U maps at 85~cm, as also noted by \citet{Bernardi09}. Finally, we note the similarity of our model with the one provided by \citet{Bernardi09}, who consider the presence of three separated components: a synchrotron-emitting plus a separated Faraday-rotating components beyond the ``bubble'', which correspond to our spatially extended, synchrotron-emitting and Faraday-rotating ``ring'' component, which surrounds the ``bubble''. The main differences from their model are the absence of the very nearby ``curtain'' component and the position of the ``bubble'' with respect to the ``ring'' component.

\section{Summary and conclusion}

We used the RM-synthesis technique on WSRT radio polarization data at $\sim 2$~m to reveal a complex Galactic synchrotron-emission foreground in a field in the Fan Region centred at $(l,b) =(137^{\circ},7^{\circ})$. We detected polarized signals with Faraday depth between -13 and +5~rad~m$^{-2}$, which can be separated into three distinct components around -5, -2, and +2~rad~m$^{-2}$, based on morphological consistency and coherence in Faraday depth space. For the first time, cross-correlation was applied to identify and characterize these polarized structures in Faraday depth space. The low Faraday depth values of all components suggest nearby locations of the emission.  

The structure at [0,+5]~rad~m$^{-2}$ (called the ``curtain'') very likely corresponds to enhanced synchrotron emission due to compressed magnetic fields in the Local Bubble wall at around 70~pc distance, in agreement with the location of the wall from optical polarized starlight data. The low but positive Faraday depth of this component suggests a parallel magnetic field component directed towards the observer.

The second component centred at -1~rad~m$^{-2}$ (called the ``ring'' component) is located just behind the Local Bubble. It is expected to be spatially extended and Faraday-thick, and therefore only partially detected in these data.

The third component (the ``bubble'') is located within the extended ``ring''. Its regular circular shape suggests relatively uniform electron density and magnetic field structure, as is expected within the Fan region. The Faraday depth of the ``bubble'' in combination of its non-detection in H$\alpha$ indicates that it cannot be created by density enhancement alone, but needs to have enhanced magnetic field strength as well, contrary to earlier detected regular polarization structures  \citep{Gray98,Uyaniker02,Uyaniker03}. Several possible associations of the bubble with ISM objects we discussed, but the most likely explanation is the presence of a nearby ($\sim200$~pc) relic Str\"{o}mgren sphere, associated with old unidentified white dwarf star and expanding in a low-density environment.

Polarized intensity from both the ``ring'' and the ``bubble'' components show a power spectrum that has a power law, which is indicative of turbulence. Also, the discrete Faraday depth components suggest discrete, small-scale synchrotron emitting structures in the ISM, as noted earlier by \citet{Brentjens11}.

This analysis shows that radio-polarimetry can detect magnetized objects in the ISM that is not detectable in any other means. If fainter and Faraday thin structures lie along the line of sight toward the Fan region, this will be seen by deeper low-frequency observations. In the near future, the higher sensitivity and resolution of modern low-frequency arrays, such as the LOw Frequency ARray \citep[LOFAR,][]{Heald11} may reveal these new components. The study of these spatially very extended and polarized foreground components will benefit from accurate Faraday rotation measurements (e.g. to separate the low $\phi$ ``curtain'' and ``ring'' components), as well as from the wide-field and accurate polarimetric imaging capabilities. A LOFAR observation over its full bandwidth (about 200~MHz from LBA to HBA high) will improve the RM-synthesis maximum scale sensitivity in Faraday depth space, allowing to be detected also the eventually Faraday thick components along the line of sight. The low-frequency arrays of next generation will sensitively probe weak magnetic fields with low Faraday depths, which may reveal more (``bubble''-like) magnetized objects.

\paragraph*{\textit{Acknowledgments}}
The Westerbork Synthesis Radio Telescope is operated by the Netherlands Foundation for Research in Astronomy (ASTRON) with financial support from the Netherlands Organization for Scientific Research (NWO). The research leading to these results has received funding from the European Union's Seventh Framework Programme (FP7/2007-2013) under grant agreement number 239490. This work is part of the research programme 639.042.915, which is (partly) financed by the Netherlands Organisation for Scientific Research (NWO). This research has made use of the SIMBAD database, operated at the CDS, Strasbourg, France. We thank the anonymous referee for carefully reading the manuscript and providing helpful comments and suggestions in the preparation of the final manuscript.

\end{document}